\documentclass[twocolumn]{aa}  

\usepackage{graphicx}
\usepackage{txfonts}
\usepackage{xspace}
\usepackage{stackrel, amsmath, amssymb}

\newcommand{\kms}{km$\,$s$^{-1}$\xspace}
\newcommand{\nnh}{N$_2$H$^+$\xspace}
\newcommand{\nnd}{N$_2$D$^+$\xspace}
\newcommand{\nqn}{N$^{15}$NH$^+$\xspace}
\newcommand{\qnn}{$^{15}$NNH$^+$\xspace}
\newcommand{\ratio}{$^{14}$N/$^{15}$N\xspace}
\newcommand{\ratioa}[1]{$\rm ^{14}N/^{15}N= #1$\xspace}

\newcommand{\tmb}{$T_\mathrm{MB}$\xspace}
\newcommand{\ncol}{$N\rm _{col}(N_2H^+)$\xspace}
\newcommand{\ncolq}{$N\rm _{col}(N^{15}NH^+)$\xspace}
\newcommand{\tex}{$T_\mathrm{ex}$\xspace}
\newcommand{\tdust}{$T_\mathrm{dust}$\xspace}

\newcommand{\nrare}{$\rm ^{15}N$\xspace}
\newcommand{\am}{$\rm NH_3$\xspace}
\newcommand{\fco}{$f_\mathrm{D}$\xspace}

\begin{document}

   \title{First sample of  \nnh nitrogen isotopic ratio measurements in low-mass protostars\thanks{This work is based on observations carried out with the IRAM 30 m Telescope. IRAM is supported by INSU/CNRS (France), MPG (Germany) and IGN (Spain).}}

   \author{E. Redaelli
          \inst{1}
          \and
         L. Bizzocchi \inst{1} \and
                      P. Caselli \inst{1} }
           
  \institute{Centre for Astrochemical Studies, Max-Planck-Institut f\"ur extraterrestrische
              Physik, Gie\ss enbachstra\ss e 1, D-85749 Garching bei M\"unchen (Germany) \\
              \email{eredaelli@mpe.mpg.de}
              }

  \titlerunning{\ratio ratio in YSOs}


 
  \abstract
   {The nitrogen isotopic ratio is considered an important diagnostic tool of the star formation process, and \nnh is particularly important because it is directly linked to molecular nitrogen $\rm N_2$. However, theoretical models still lack to provide an exhaustive explanation for the observed \ratio values.}
   {Recent theoretical works suggest that the \ratio behaviour is dominated by two competing reactions that destroy \nnh: dissociative recombination and reaction with CO. When CO is depleted from the gas phase, if \nnh recombination rate is lower with respect to the \nqn one, the rarer isotopologue is destroyed faster. {In prestellar cores, due to a combination of low temperatures and high densities, most CO is frozen in ices onto the dust grains, leading to high levels of depletion. On the contrary, in protostellar cores, where temperature are higher, CO ices evaporate back to the gas phase}. This implies that the \nnh isotopic ratio in {protostellar cores} should be lower than the one in prestellar cores, and consistent with the elemental value of $\approx 440$. We aim to test this hypothesis, producing the first sample of $ \text{\nnh}/ \text{\nqn}$ measurements in low mass protostars.}
   {We observe the \nnh and \nqn {lowest} rotational transition towards six young stellar objects in Perseus and Taurus molecular clouds. We model the spectra with a custom \texttt{python} code using a constant \tex approach to fit the observations. We discuss in appendix the validity of this hypothesis. The derived column densities are used to compute the nitrogen isotopic ratios.}
   {Our analysis yields an average of $\text{\ratio}|_\mathrm{pro} = 420\pm 15$ in the protostellar sample. This is consistent with the protosolar value of $440$, and significantly lower than the average value previously obtained in a sample of prestellar objects. }
   {Our results are in agreement with the hypothesis that, when CO is depleted from the gas-phase, dissociative recombinations with free electrons destroy \nqn faster than \nnh, leading to high isotopic ratios in prestellar cores, where carbon monoxide is frozen onto dust grains.}


   \maketitle
%

\section{Introduction}
Nitrogen is the fifth element in the universe for abundance. In molecular gas, it is believed that its main reservoir is $\rm N_2$. However this molecule, similarly to $\rm H_2$, is not directly observable in cold environments due to the lack of permanent dipole moment. Diazenylium (\nnh) forms directly from molecular nitrogen through reaction with $\rm H_3^+$, and it emits bright lines at millimetre wavelengths. It is therefore traditionally considered a good probe of the bulk of nitrogen gas in molecular clouds. Furthermore, nitrogen bearing species are less affected by depletion at very high densities with respect to C- and O-bearing ones, and hence nitrogen hydrides are usually used to trace the innermost parts of star forming regions. Recent works \citep{Caselli02b, Bergin02, Pagani07, Redaelli19} showed that in fact also \nnh and the deuterated \nnd experience freeze-out onto dust grains in very dense gas, but to a much lesser extent with respect to CO, $\rm HCO^+$ and isotopologues. \par
Nitrogen is present in two stable isotopes, the main $\rm ^{14}N$ and the much less abundant $\rm ^{15}N$. In the last decades, its isotopic ratio \ratio has been extensively studied both in the Solar System and in the interstellar medium (ISM), since it allows to link the early phases of star formation with more evolved stellar systems like our own, where pristine material is still preserved in icy bodies (e.g \citealt{Altwegg19}). In fact, unlikely for instance oxygen isotopic ratios, the nitrogen one is spread over a wide range. In the primitive Solar Nebula, in situ measurements in the Solar wind find \ratioa{440-450} \citep{Marty11}, and a similar value is reported also in the Jupiter atmosphere \citep{Fouchet04}. Molecular nitrogen in the Earth's atmosphere is on the contrary enriched in \nrare (\ratioa{272}, \citealt{Nier50}). Hotspots in carbonaceous chondrites, which are believed to represent some of the most pristine materials in the Solar System, can present isotopic ratios as low as 50 \citep{Bonal10}. These results led to the idea that at the moment of the birth of our planetary system, multiple nitrogen reservoir were present (see, for instance, \citealt{HilyBlant17}). \par
Measurements of \ratio in the ISM also yielded a variety of results, depending both on the environments and on the tracer. Usually nitriles (HCN, HNC, and CN) appear enriched in \nrare with respect to the protosolar value: \ratioa{160-460} were found in a sample of low-mass protostars by \cite{Wampfler14}; in prestellar cores, \cite{Hily-Blant13a, Hily-Blant13b} found values of $140-360$ in HCN/HNC and $470-510$ in CN. In high-mass star forming regions, \cite{Colzi18a} found \ratioa{250-650}. It is important to highlight, however, that all these measurements were derived using the double-isotope methods, i.e. observing the $^{13} \rm C$-bearing species instead of the more abundant and optically thick $^{12} \rm C$-bearing ones. As a consequence, these results depend on the assumed carbon isotopic ratio $\rm ^{12}C / ^{13}C$, and they can be up to a factor of 2 off, as recently shown by \cite{Colzi20}. \ratio observations with diazenylium, on the contrary, do not depend on assumptions of this kind, but are made more difficult by a) the intrinsic weakness of the \nqn and \qnn lines, requiring long integration times, especially for low-mass star forming regions; b) the hyperfine anomalies shown by the \nnh (1-0) line, in particular in cold and dense environments. As a consequence, studies of the N-fractionation in \nnh are rarer. \cite{Daniel13} found $400-600$ towards Barnard 1b, which hosts two very young protostellar objects \citep{Gerin15}. In OMC-2, a protocluster containing several protostars, \cite{Kahane18} derived \ratioa{190-380}. \cite{Bizzocchi13} derived extremely high levels of \nrare depletion in the prestellar core L1544. This result was later confirmed by \cite{Redaelli18}, who found $\text{\ratio} = {580-1000}$ in a small sample of prestellar sources. In the high mass regime, \cite{Fontani15} found \ratioa{180-1300}. \par
From the theoretical point of view, it is currently difficult to interpret all these observational results. Chemical models are often able to reproduce the \nrare enrichment in nitriles with respect to the elemental value (assumed to be equal to the protosolar value of $440$, see e.g. \citealt{Roueff15}), even though more recent results seem to be in disagreement \citep{Wirstrom17}. On the contrary, the case of the high \ratio values shown in \nnh is still puzzling.   \par
In the last year, however, two possible solutions to give explanation to the \nnh fractionation observations have been proposed. \cite{Furuya18} suggested that the depletion in \nrare is inherited from the initial stages of the core evolution, when the gas density is low enough that UV photons can penetrate. Molecular nitrogen is selectively photodissociated, since the rare $\rm N^{15}N$ is not abundant enough for self-shielding. This leads to a $^{15}$N enrichment in the atomic nitrogen (N) gas. When N freezes out onto dust grains, where it is rapidly transformed in \am ices, the bulk gas results depleted in heavy nitrogen, while the \am ices are enriched. The weak point of this theory is that the selective photodissociation works at low-to-moderate visual extinction ($A_\mathrm{V} \lesssim 1.5\, \rm mag$), when ices cannot efficiently form,  so that only a small fraction of \nrare can be effectively trapped in ices. \par
Another possible explanation has been proposed by \cite{Loison19}, who used the 3-phase chemical model Nautilus \citep{Ruaud16} to follow the nitrogen chemistry during the star formation process. They suggest that the  $^{15}$N-antifractionation seen in \nnh can be due to a difference in the dissociative recombination (DR) rates for the different isotopologues. The main \nnh destruction pathways, according to their model, are the following: 
\begin{align}
\mathrm{N_2H^+ + CO} & \rightarrow \mathrm{HCO^+ +N_2}  \; ,\\
\mathrm{N_2H^+ + e} & \rightarrow \mathrm{N_2 + H} \; .
\end{align}
When gas-phase CO abundance is low, such as in cold and dense prestellar cores, where CO is mainly frozen onto dust grain surfaces, the dominant reaction is the DR one. If the DR rate ($\kappa_\mathrm{DR}$) of \nnh is lower than those of \nqn and \qnn, the \nnh/\nqn and \nnh/\qnn ratios can be significantly larger than the elemental value. When on the other hand CO abundance is high, its reaction with \nnh becomes the dominant one and the molecular isotopic ratio decreases back to the elemental value. Recent laboratory work \citep{Lawson11} showed that \nnh isotopologues exhibit DR rates that vary up to 20\% in value, a discrepancy which is of the same order of magnitude of the one {hypothesised} by \cite{Loison19}, who assumed {a DR rate of \nnh 50\% lower than that of \nqn}. The laboratory results are in the direction opposite to the one required by the Loison's theory (i.e. the DR rate of \nnh is higher than that of \nqn). However, no data are available at the ISM low temperatures, since the experiments were performed at room temperature. {More recently, \cite{Hily-Blant20} investigated further this topic. They model the nitrogen isotopic fractionation of several species during the collapse of a core. The novelty of this study is that the chemistry and the dynamics are run simultaneously, whilst in most other works the chemical evolution is simulated in a quasi-static fashion. Their results show that the isotopic dependence of the adsorption rates plays an important role on the evolution of the nitrogen fractionation during the collapse, but the model still fails in reproducing the \ratio values observed in \nnh. The paper concludes, in agreement with \cite{Loison19}, that different DR rates for the distinct diazenylium isotopologues could explain the observed values. In particular, they find that $\kappa_\mathrm{DR} (\text{\nnh}) = 2-3 \times \kappa_\mathrm{DR} (\text{\nqn})$ is needed to reproduce the observations. } \par
There are a few observational hints that point towards this direction: for instance, the \ratio values measured in OMC-2 and Barnard 1b, which hosts YSOs and are therefore warmer, are lower than the ones measured in the prestellar cores sample of \cite{Redaelli18}. In the high mass regime, \cite{Fontani15} found an anti-correlation between \nnh/\nqn and \nnh/\nnd. Since the deuteration process is highly favoured by the CO depletion \citep{Dalgarno84}, these results also suggest that \nnh/\nqn is larger where CO is mostly absent from the gas-phase. \par
In order to test the hypothesis of \cite{Loison19}, we performed observations of \nnh and \nqn (1-0) lines towards six young stellar objects, hence obtaining the first sample of \ratio in diazenylium in low-mass {prostostellar cores}. If the theory is correct, we expect  a lower \ratio with respect to prestellar {cores}.

\section{Observations \label{sec:observatios}}
The observations of \nnh and \nqn were performed with the Institut de Radioastronomie Millim\'etrique (IRAM) $30\,$m telescope, located at Pico Veleta (Spain), during two different sessions (August 2019 and December 2019). The weather was good for 3mm observations during the summer ($0.30 < \tau_{225\mathrm{GHz}}<0.70$), and very good in winter ($ \tau_{225\mathrm{GHz}}<0.30$). The pointing was frequently checked on bright nearby sources, and found to be usually accurate within $5''$. Mercury and Uranus were used as focus calibrators.  We used the EMIR E0 frontend, in two different frequency setup, the first centred on the \nnh (1-0) frequency ($93173.3991\, \rm MHz$) and the second centred on the \nqn (1-0) frequency ($ 91205.6953\, \rm MHz$). EMIR was combined with the VESPA backend, set to high spectral resolution ($\Delta \nu = 20 \rm \, kHz$, corresponding to $0.06 \,$\kms at $93\, \rm GHz$). The beam size at $93\, \rm GHz$ is $\theta_\mathrm{beam} = 27''$. \par
\begin{table*}[t]
\renewcommand{\arraystretch}{1.4}
\centering
\caption{Summary of the targeted sample. All data are from \cite{Emprechtinger09}. \label{sources}}

\begin{tabular}{ccccc}
\hline
Sources        & RA (J2000) & Dec (J2000) & $T_\mathrm{dust}$ & $f_\mathrm{D}$\tablefootmark{a}  \\ 
       &   hh:mm:ss.s  &    dd:mm:ss  &  K & \\
\hline
IRAS 03282      & 03:31:21.0 & 30:45:30    & $23\pm 2 $        &  $3.9\pm 0.9$    \\
L1448 IRS2     & 03:25:22.4 & 30:45:12    & $27 \pm 2$        &  $2.6 \pm0.6$                                  \\
L1448 C        & 03:25:38.8 & 30:44:05    & $32  \pm 2 $         &     $2.8  \pm0.6  $     \\
L1455 A1           & 03:27:42.1 & 30:12:43   &$ 37\pm 2 $       &  $1.5 \pm0.4$ \\
Barnard 5 IRS1 & 03:47:41.6 & 32:51:42    & $50 \pm 2$       &   $0.30 \pm0.09$    \\ 
L1527 & 04:39:53.5  & 26:03:05 & $27\pm 2$ &  $2.3 \pm0.5$ \\

\hline
\end{tabular}
\tablefoot{
\tablefoottext{a}{This is computed as the ratio between the canonical CO abundance and the observed one, the former being obtained from $\rm H_2$ column density via $X_\mathrm{mol}(\mathrm{C^{18}O})/X_\mathrm{mol}(\mathrm{H_2}) = 1.7\times 10^{-7}$ \citep{Frerking82}}.
}
\end{table*}

\begin{table*}[!t]
\renewcommand{\arraystretch}{1.4}
\centering
\caption{The obtained best-fit parameters for \nnh and \nqn in the source sample, and the derived nitrogen isotopic ratio. \label{para_results}}
\begin{tabular}{llccccc}
\hline
Source               & Molecule & $N_\mathrm{col}$              & \tex    & $V_\mathrm{lsr}$ & $FWHM$  & \ratio\\
 &          & $10^{10} \rm \, cm^{-2}$ & $\rm K$ & \kms             & \kms  &  \\
 \hline
 
IRAS 03282           & \nnh     &   $(1.22 \pm 0.02)\times 10^3$        & $7.22 \pm 0.13$   &    $6.9607 \pm 0.0014$    &    $0.387 \pm 0.003$  &   \\
                     & \nqn     &      $3.6 \pm 0.3  $    &     7.22\tablefootmark{a}   &           $6.937\pm 0.011$       &   $0.31 \pm 0.03$  & $340\pm30$   \\
L1448 IRS2           & \nnh     &        $(2.76 \pm 0.04)\times 10^3$    &  $6.00 \pm 0.03$  &     $4.1082 \pm 0.0016 $  &    $0.506 \pm 0.003$  &   \\
                     & \nqn     &      $4.8 \pm 0.3$           &  6.00\tablefootmark{a}  &     $4.109 \pm 0.012 $   &   $0.45 \pm 0.03$  & $580\pm40$   \\
L1448 C               & \nnh     &    $(2.54 \pm 0.07)\times 10^3$ &  $6.41\pm 0.09  $&  $ 5.073 \pm 0.005$ &  $   0.814 \pm 0.013 $  & \\
                     & \nqn     &         $5.9 \pm 0.4$    &     $6.41$\tablefootmark{a}     &      $5.05\pm 0.03$    &   $0.97\pm 0.07$ & $430\pm30$     \\
L1455 A1             & \nnh     & $(1.13 \pm 0.04)\times 10^3$  &$  7.2 \pm 0.4$ &    $4.987 \pm 0.005$   &   $ 0.645\pm 0.011$  & \\
                     & \nqn     &      $2.0\pm 0.2  $    &  $7.2$\tablefootmark{a}  &  $5.05\pm 0.03$  &   $0.52 \pm 0.07$  & $570\pm60$ \\
Barnard 5            & \nnh     &     $(1.44 \pm 0.03)\times 10^3$     &  $6.32 \pm 0.08$  &   $10.2696 \pm 0.0017$  &  $0.449 \pm 0.004$  & \\
                     & \nqn     &    $ 3.9 \pm 0.3$   &   $6.32$\tablefootmark{a}      &  $10.292 \pm 0.016$  &  $0.49 \pm 0.04$ & $370\pm30$ \\
L1527                & \nnh     & $(7.58\pm 0.18)\times 10^2 $    & $4.99 \pm 0.05$  &   $5.9056 \pm 0.0011   $ &  $   0.302 \pm 0.003 $  &  \\
                     & \nqn     &   $1.8 \pm 0.3$    & $4.99$\tablefootmark{a} &   $5.952 \pm 0.024$   &  $0.29 \pm 0.06$  & $420\pm70$    \\
                     \hline
\end{tabular}
\tablefoot{
\tablefoottext{a}{The excitation temperature for \nqn (1-0) is fixed to the value derived from the corresponding \nnh fit.}
}

\end{table*}

The source sample consists of six young stellar objects (YSOs), five of which belong to the Perseus molecular cloud. The sixth one (L1527) is located in the Taurus cloud. The {protostellar cores} were selected from the sample of \cite{Emprechtinger09}, who investigated the deuterium fractionation of \nnh in YSOs. The choice was first led by bright emission in the \nnh (1-0) transition, maximising the chance of detecting the much weaker \nqn line. Furthermore, we tried to select objects with different recorded dust temperature and CO depletion factor, since these two parameters may play a role in the nitrogen fractionation (see Introduction). Finally, in order to minimise environmental effects, the initial sample consisted of cores from a single molecular cloud. However, due to its high elevation from Pico Veleta, Perseus was not continuously observable, and hence L1527 was added as a filler source. At the cloud distances ---$295\, \rm pc$ for Perseus and $135\, \rm pc$ for Taurus \citep{Zucker18,Schlafly14}--- the telescope angular resolution corresponds to $0.04$ and $0.02\, \rm pc$, respectively. \par
The integration times were $23\, \rm min$ for \nnh and between 3 and $8\,\rm h$ for \nqn, resulting in $rms = 20-25\, \rm mK$ (main isotopologue) and $rms = 4-6\, \rm mK$ (rare isotopologue). Table \ref{sources} summarises the source sample, including coordinates, dust temperatures ($T_\mathrm{dust}$), and CO depletion factors ($f_\mathrm{D}$). \par
The data were reduced using GILDAS/CLASS package\footnote{Available at http://www.iram.fr/IRAMFR/GILDAS/.}, and they were calibrated into main beam temperature ($T_\mathrm{MB}$) using the tabulated beam efficiency ($\eta_\mathrm{MB} = 0.80$) and forward efficiency ($F_\mathrm{eff} = 0.95$).

\section{Results}
\begin{figure*}
   \centering
  \includegraphics[width=0.82 \textwidth]{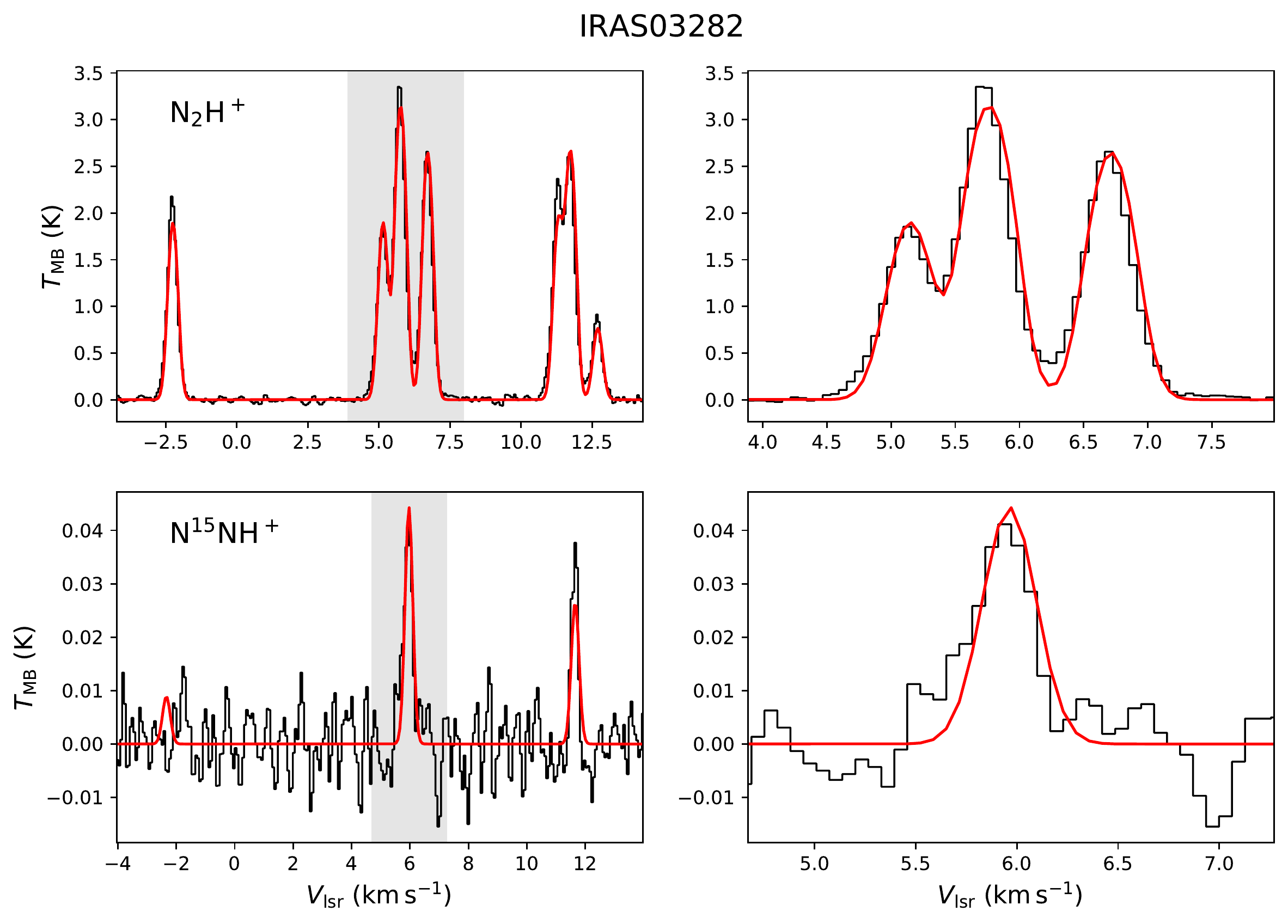}
      \caption{Observed spectra (black histograms) and best-fit models (red curves) in IRAS 03282. The top panels refer to \nnh, the lower ones to \nqn. The right panels show the zoom-in of the grey-shaded area in the corresponding left plots.    \label{spec_03282}}
   \end{figure*}

\begin{figure*}[!h]
   \centering
  \includegraphics[width=0.82 \textwidth]{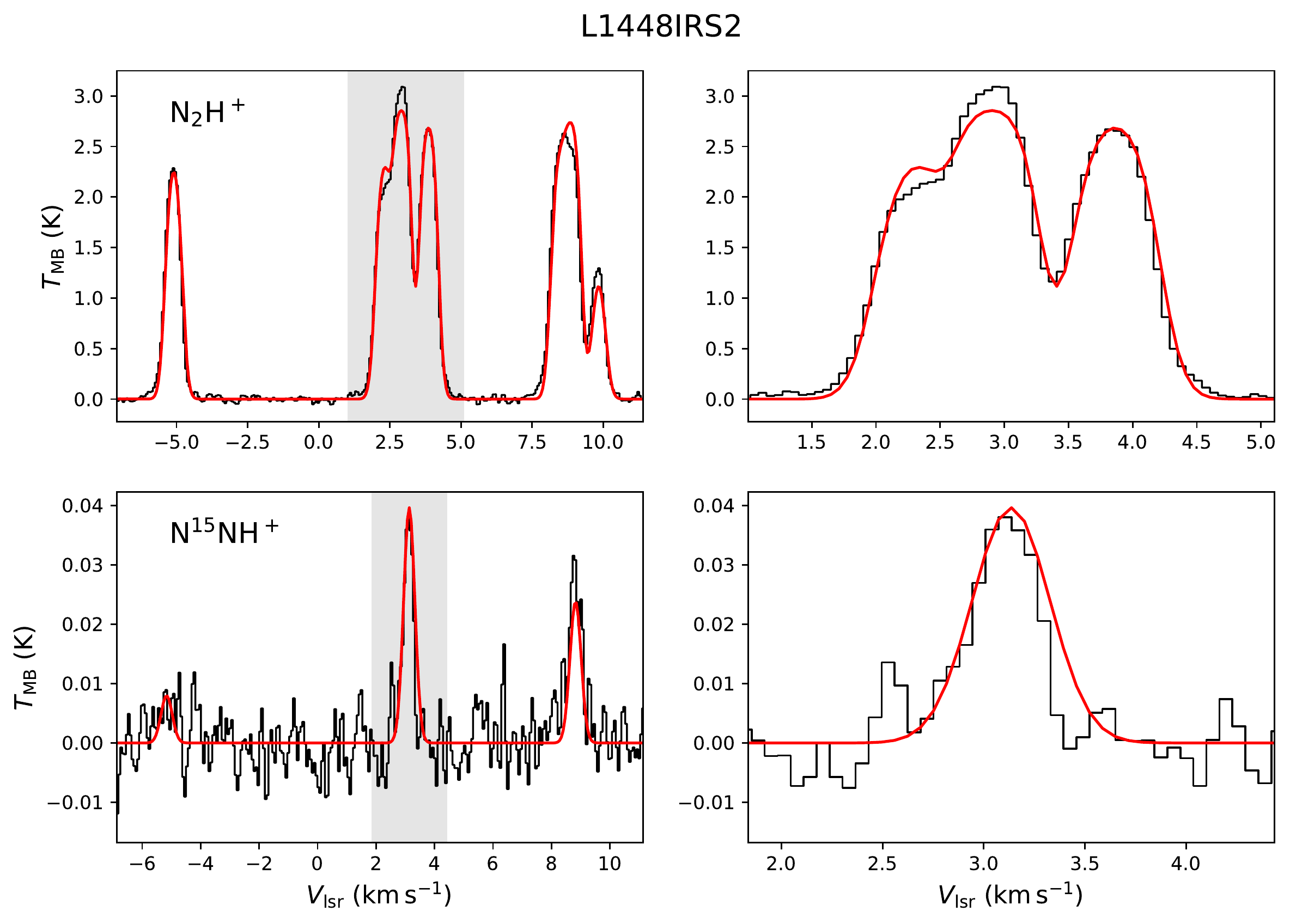}
      \caption{Same as in Figure \ref{spec_03282}, but for L1448 IRS2.         \label{spec_IRS2}}
   \end{figure*}

\begin{figure*}[!h]
   \centering
  \includegraphics[width=0.82 \textwidth]{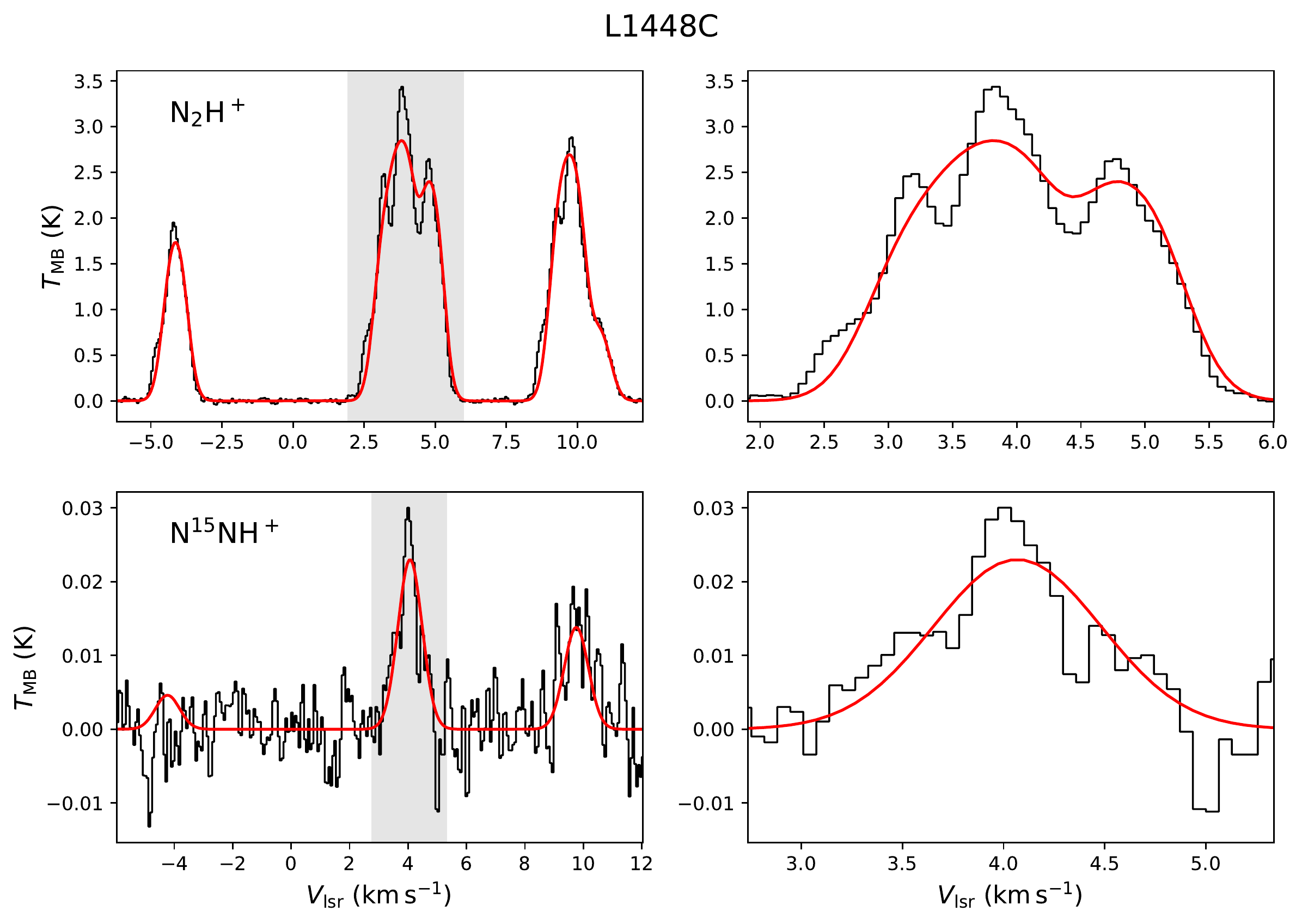}
      \caption{Same as in Figure \ref{spec_03282}, but for L1448 C  \label{spec_1448C}}
   \end{figure*}

\begin{figure*}[!h]
   \centering
  \includegraphics[width=0.82 \textwidth]{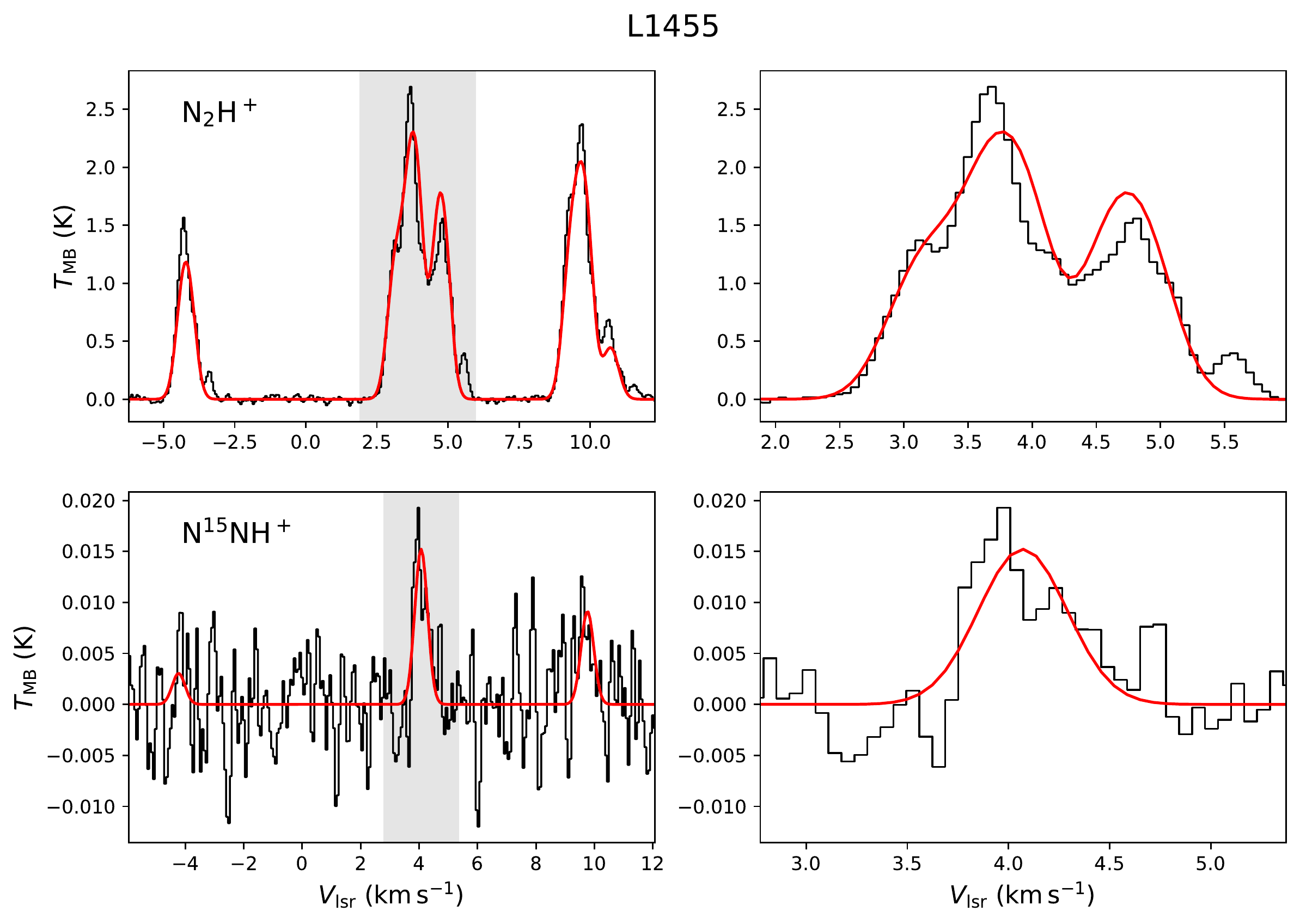}
      \caption{Same as in Figure \ref{spec_03282}, but for L1455 A1.  \label{spec_1455}}
   \end{figure*}

\begin{figure*}[!h]
   \centering
  \includegraphics[width=0.82 \textwidth]{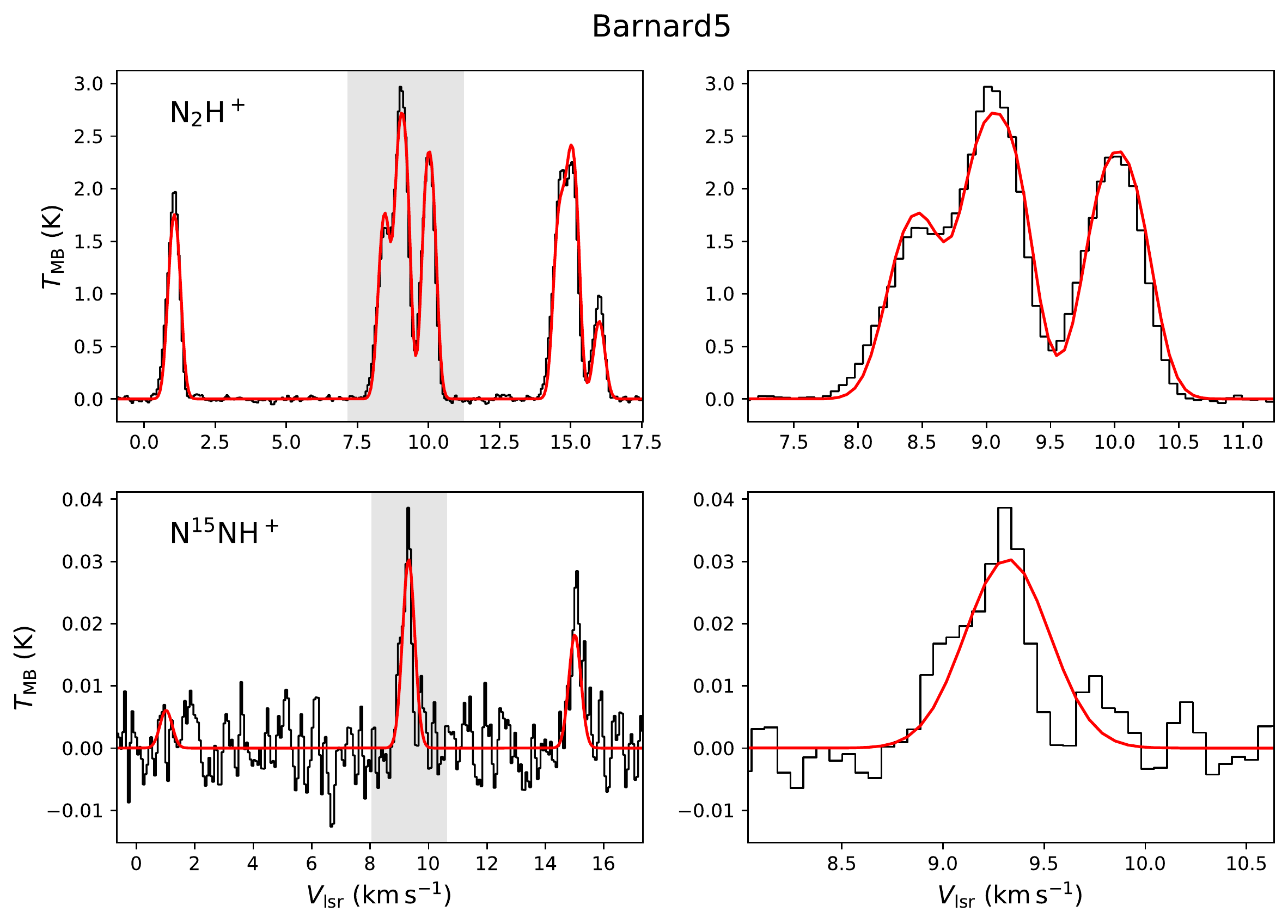}
      \caption{Same as in Figure \ref{spec_03282}, but for Barnard 5. \label{spec_B5}}
   \end{figure*}

\begin{figure*}[!h]
   \centering
  \includegraphics[width=0.8 \textwidth]{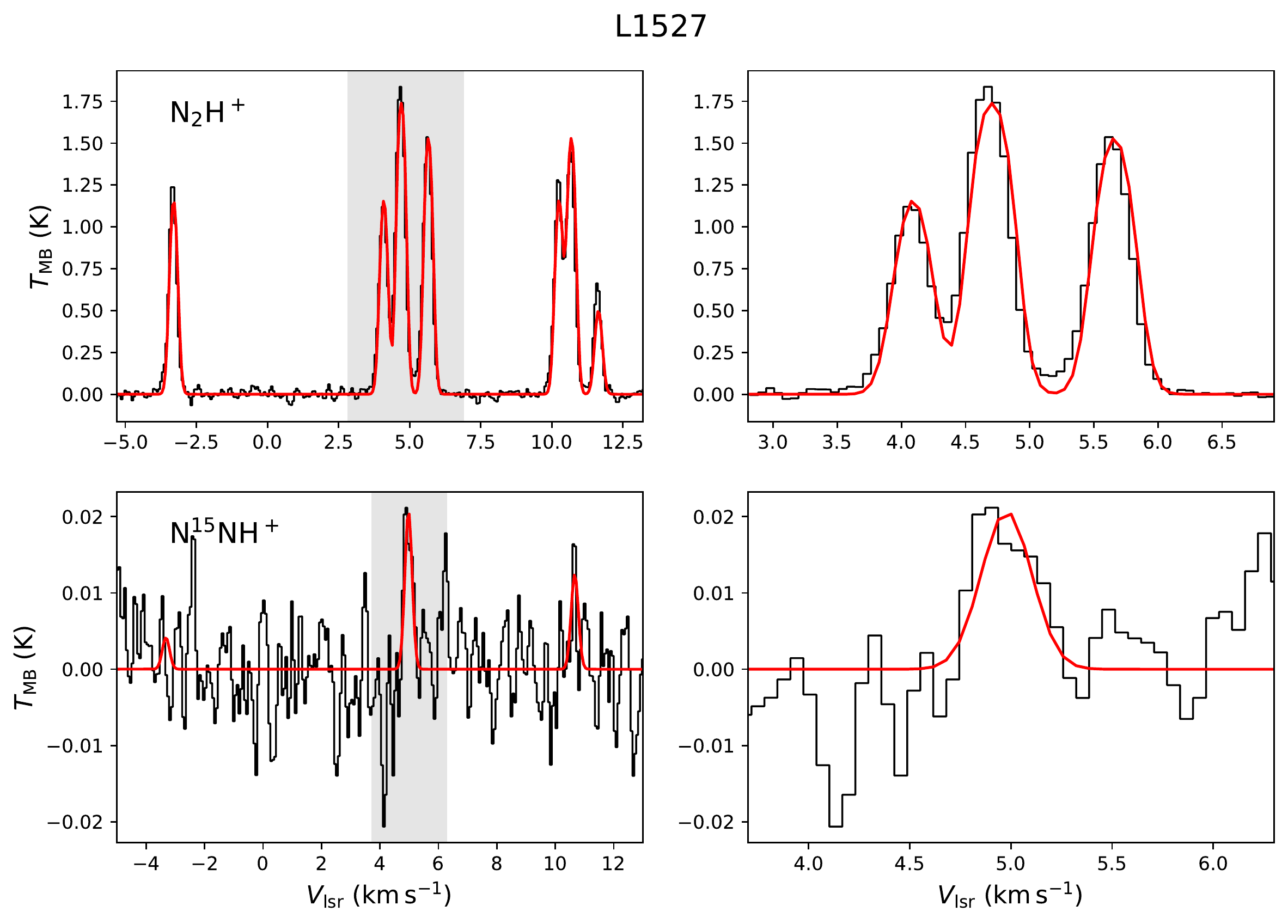}
      \caption{Same as in Figure \ref{spec_03282}, but for L1527.  \label{spec_1527}}
   \end{figure*}

The observed spectra are shown in Figures \ref{spec_03282} to \ref{spec_1527} as black histograms. The top panels present the main isotopologue, whilst the bottom panels show the \nqn transition. The \nnh (1-0) is well detected in all sources, with the brightest hyperfine component presenting \tmb values higher than $2.5\,$K (with the exception of L1527, where the central component reaches $1.75\,$K).  The $rms$ of these observations hence translates in a signal-to-noise ratio $SNR>75$. The \nqn (1-0) line is much weaker, which justifies our choice of selecting only YSOs with previously detected bright \nnh emission. The peak brightness \tmb ranges from $20$ to $40\, \rm mK$. The transition is however detected in all sources, with $SNR$ going from $3$ (L1527) to 	$SNR = 8$ (Barnard 5).

\section{Analysis\label{sec:analysis}}
The goal of this work is to derive the nitrogen isotopic ratio of \nnh in each source. In the assumption that the two isotopologues are co-spatial, \ratio is computed as the ratio of column densities \ncol /\ncolq. Reliable estimates of the latter are therefore needed. In order to estimate $N_\mathrm{col}$, we use a custom \texttt{python} code that implements a constant \tex ($C\_T_\mathrm{ex}$) fit of the hyperfine structure (see also \citealt{Melosso20}). The code works in a similar fashion as the HFS routine of the CLASS package, but the fit parameters are the molecular column density $N_\mathrm{col}$ and the excitation temperature \tex (instead of the optical depth $\tau$), together with the kinematic parameters (centroid velocity $V_\mathrm{lsr}$ and line full width at half maximum $FWHM$). Furthermore, it allows easily to fit multiple velocity components and to set a subset of the parameters as constrains. Accurate frequencies of the individual hyperfine components are taken from \cite{Dore09} for \nqn and from our own calculations, based on the results from \cite{Cazzoli12}, for \nnh. The values of the partition functions come from our own calculations.


\subsection{Single velocity component analysis\label{sec:SingleVel}}
The first implemented fitting strategy is based on a single velocity component analysis. For each source, we first analyse the \nnh (1-0) line, which {has a total optical depth $\tau_\mathrm{tot}>> 1$ and is therefore} optically thick, {hence allowing} to derive \ncol and \tex at the same time, using a single velocity component. The weaker \nqn (1-0) transition is optically {thinner} {($\tau_\mathrm{tot} << 1$)}, and therefore \ncolq and \tex are degenerate parameters that cannot be fitted simultaneously. In this case, then, we fix \tex to the value derived in the corresponding \nnh analysis. Table \ref{para_results} summarises the best-fit parameters. The computed models are also shown in Figure \ref{spec_03282}-\ref{spec_1527} as red curves overlaid to the observations. For most sources, the fit is reasonably good, and the model is able to correctly reproduce the linewidths and the intensity ratios of the different hyperfine components. There are two protostars, however, for which the fit is not optimal, as one can see especially from the zoom-in on the central triplet of the \nnh (1-0) transition. L1448 C (figure \ref{spec_1448C}, top panels) presents a broad wing on one side of the spectrum, which is not reproduced by a single velocity component model. Furthermore, the model fails also to reproduce the hyperfine flux ratios. The situation is similar for L1455 A1 (figure \ref{spec_1455}, top panels), where not only the intensities of the components are not well matched, but there is an evident feature on the red side of the line, which appears as a second, isolated velocity component. This was not detected by \cite{Emprechtinger09} due to the limited sensitivity of the observations. We discuss in Sec. \ref{sec:MultiVel} the results of a multiple velocity component fit for these two objects. \par
\begin{table*}[!t]
\renewcommand{\arraystretch}{1.4}
\centering
\caption{Best-fit values obtained with the two velocity component fit in L1448 C for both \nnh and \nqn (1-0) transitions. The last column reports the total column density of each isotopologue, from which \ratio is computed.\label{tab:L1448C_2}}
\begin{tabular}{cccc}
\hline
Parameter                              & Comp. 1 & Comp. 2 & 1+2 \\
\hline
\multicolumn{4}{c}{\nnh}                                         \\
$N_\mathrm{col}/10^{13}\, \rm cm^{-2}$ &    $0.52 \pm 0.06$    &  $2.75 \pm 0.09$        &  $3.27\pm 0.11$   \\
$T_\mathrm{ex}/K$                      &    6.41\tablefootmark{a}     & $4.76 \pm 0.13$&     \\
$V_\mathrm{lsr}/\text{\kms}$           & $5.124 \pm 0.006$        &    $5.025 \pm 0.008 $    &     \\
$FWHM/\text{\kms}$                     &   $ 0.49 \pm 0.02 $    &    $0.99 \pm 0.02$     &     \\
\hline
\multicolumn{4}{c}{\nqn}                                         \\
$N_\mathrm{col}/10^{10}\, \rm cm^{-2}$ &   $0.5 \pm 0.4$      & $6.4\pm0.7$& $6.9 \pm 0.8$     \\
$T_\mathrm{ex}/K$                      &    6.41\tablefootmark{b}     &    4.76\tablefootmark{b}      &     \\
$V_\mathrm{lsr}/\text{\kms}$           &   5.124\tablefootmark{b}      &     5.025\tablefootmark{b}    &     \\
$FWHM/\text{\kms}$                     &     0.49\tablefootmark{b}    &     0.99\tablefootmark{b}    &    \\
\hline
\end{tabular}
\tablefoot{
\tablefoottext{a}{Parameter fixed to the best-fit value found with the one-component analysis.} \\
\tablefoottext{b}{Parameter fixed to the corresponding one in \nnh analysis.}
}
\end{table*}
A strong assumption made in our approach is that the different hyperfine components share the same excitation temperature (constant \tex assumption), which is often debatable in the case of \nnh. In Appendix \ref{app:LTE} we discuss this hypothesis in greater detail. We demonstrate that due to a combination of broader linewidths and lower optical depths, the hyperfine anomalies are expected to be weak, and definitely less important than in prestellar cores.

\subsection{Multiple velocity component analysis\label{sec:MultiVel}}
As seen in Sec. \ref{sec:SingleVel}, the fitting routine is able to reproduce reasonably well the line profiles of the \nnh (1-0) line for four sources (IRAS 03282, L1448 IRS2, Barnard 5, and L1527). For the remaining two the obtained fits are worse, since they cannot reproduce the linewidths and the hyperfine intensities correctly. In this Section, we report the modelling of L1448 C and L1455 A1 using a multiple velocity component fit. We will show that the derived isotopic ratios are consistent within uncertainties with the simpler, one-component analysis. In the discussion (Sec. \ref{sec:Discussion}) we will therefore focus on the results from Sec. \ref{sec:SingleVel}.

\subsubsection{L1448 C}
As visible in the top panels of Figure \ref{spec_1448C}, the \nnh (1-0) line in L1448 C presents a broad wing on the blue side of the spectrum. This is most probably due to the internal kinematics of the core, which is known to power an extended bipolar outflow (see e.g. \citealt{Bachiller90}). The one-component fit routine then models a broad line (the $FWHM = 0.8 \,$\kms is the highest in the sample, see Table \ref{para_results}), but is not able to reproduce the narrower peaks of the hyperfine components, in particular in the central triplet. We therefore tried to fit the observed spectrum with two components, a narrow one and a broad one. \par
The fitting code has now in principle eight free parameters. However, since the two components are not strongly separated in velocity, it is not possible to derive the hyperfine intensity ratios for each group, independently. This information is crucial to derive simultaneously the optical depth (and hence the column density) and \tex. As a consequence, for one velocity component these two parameters result degenerate. We therefore had to fix the excitation temperature $T_\mathrm{ex, 1}$ of one of the two components. \par
The choice of which value to set for $T_\mathrm{ex, 1}$ is quite arbitrary, since we do not have other observations that can constrain this parameter. We therefore decided to fix $T_\mathrm{ex, 1}$ to the value derived with the single-component modelling, following the idea that this should be indicative of at least the average \tex of \nnh (1-0) in the source. We would like to highlight, however, that a change in \tex of $\approx 2\, \rm K$ translates in a change of $\approx 15$ in the isotopic ratio, since the excitation temperature is the same for both isotopologues. \par
\begin{figure*}[!h]
   \centering
  \includegraphics[width=0.8 \textwidth]{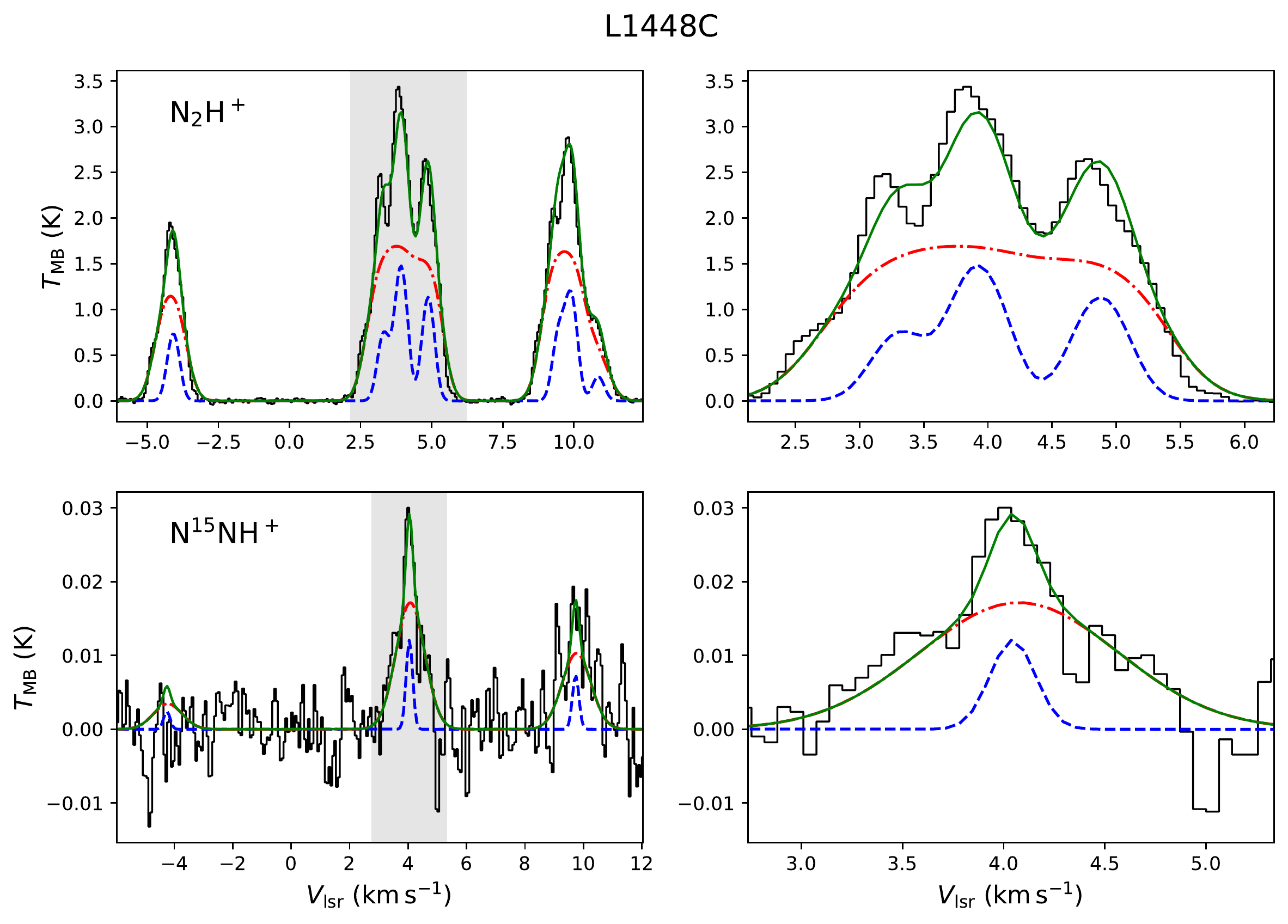}
      \caption{The obtained best fit models towards L1448 C using the two velocity component method for \nnh (top panels) and \nqn (bottom panels), overlaid to the observations (black histograms). The two components are shown with the dashed blue curve and the dashed/dotted red curve. The total modelled spectrum is shown with the solid, green curve. The right panels show the zoom-in on the central hyperfine group, highlighted with the grey shaded area in the corresponding left panels.  \label{fig:L1448C_2}}
   \end{figure*}

The code is thus run with seven free parameters. The obtained best-fit values are presented in Table \ref{tab:L1448C_2}, and the resulting fit is shown in Figure \ref{fig:L1448C_2}. The two components are separated by $\approx 0.1 \,$\kms, and the linewidth of the broad one is twice as large as the narrow one. The fit is still not ideal, but the hyperfine intensity ratios are better reproduced, as well as the aforementioned broad blue wing. In order to improve further the fit, one would need to model the kinematic structure of the core, which is beyond the scope of this work (see also comments in Appendix \ref{app:LTE}). \par
Once the \nnh (1-0) line is fitted, we can model the \nqn one with two components. The \tex values must be fixed to the \nnh values, due to low optical depth reasons (see Sec. \ref{sec:SingleVel}), so the free parameters are in principle six. However, the signal to noise ratio of the data is too low to constrain all of them, and the uncertainties on the best-fit values are $50-100$\%. Hence, in the hypothesis that the two transitions arise from the same medium, we also fixed $V_\mathrm{lsr}$ and $FWHM$ for each component to the values found from the \nnh analysis. This approach is justified by the results from the single-component analysis, which shows that these two parameters are usually consistent within $3\sigma$ uncertainties between the two isotopologues. \ncolq for each component is then the only free parameter. \par
The results are visually displayed in the bottom panels of Figure \ref{fig:L1448C_2}, and they are summarised in Table \ref{tab:L1448C_2}.  Note that the column density of the weaker component is anyhow almost unconstrained (relative uncertainty: $80$\%).
From the obtained values of column densities, we derive \ratio as the ratio of the total column densities (summing together the two components). The total column densities are significantly higher than those derived with the single-component fit. This is due to the fact that in the \nnh (1-0) line, the $T_\mathrm{ex, 2}$ of the unconstrained component (which is five times denser than the other) is low (the lowest in the sample), and lower than $T_\mathrm{ex, 1}$. The resulting \ncol is hence higher. However, the derived isotopic ratio is $\text{\ratio}|_\mathrm{2c} = 470 \pm 60$ and it is consistent within the uncertainties with the one obtained in Sec. \ref{sec:SingleVel}. 

\subsubsection{L1455 A1}
    \begin{figure*}
   \centering
  \includegraphics[width=0.8 \textwidth]{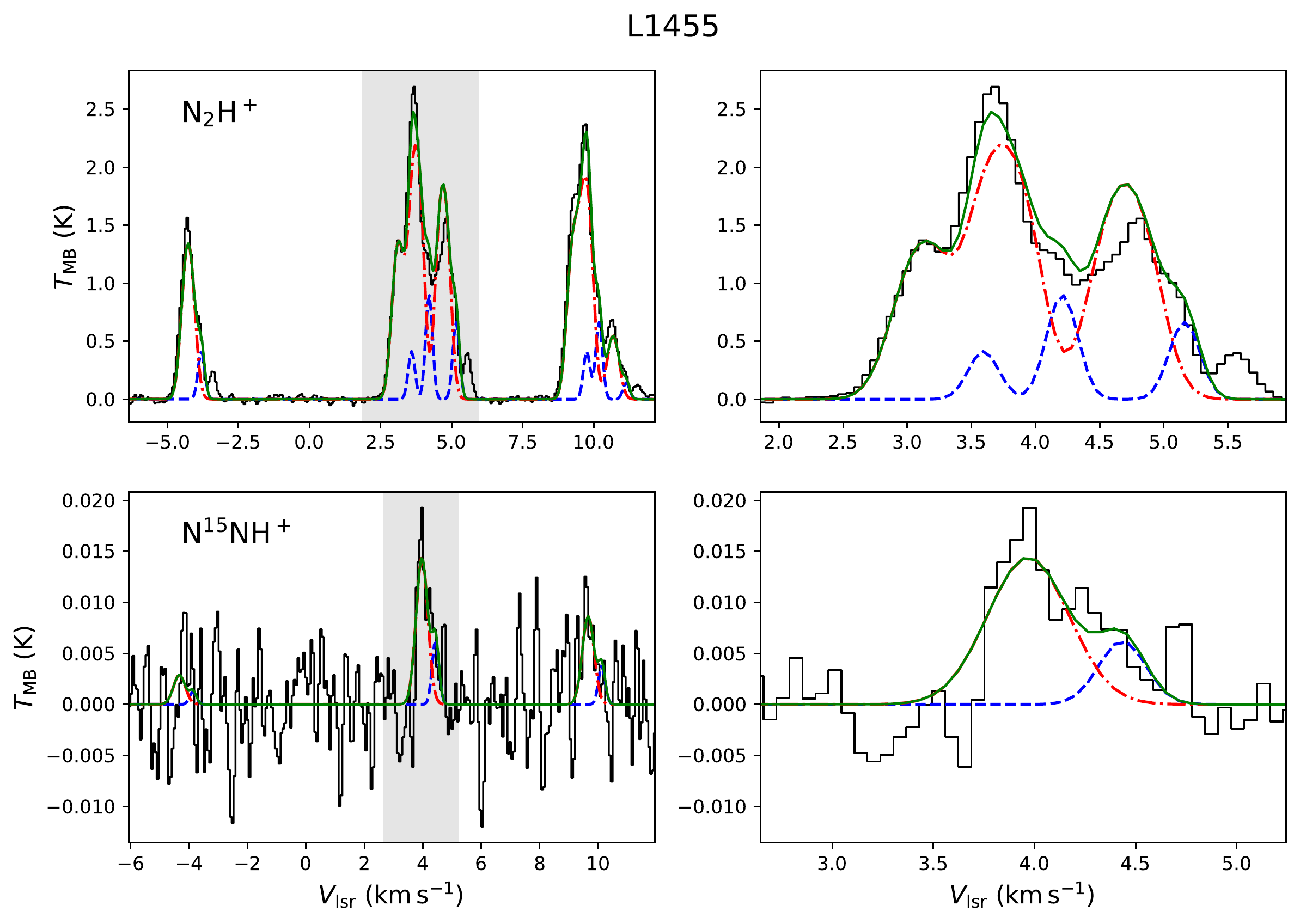}
      \caption{Same as Figure \ref{fig:L1448C_2}, but for L1455 A1. \label{fig:L1455_2}}
   \end{figure*}

\begin{table*}
\renewcommand{\arraystretch}{1.4}
\centering
\caption{Best-fit values obtained with the two velocity component fit in L1455 A1 for both \nnh and \nqn (1-0) transitions. The last column reports the total column density of each isotopologue, from which \ratio is computed.\label{tab:L1455_2}}
\begin{tabular}{cccc}
\hline
Parameter                              & Comp. 1 & Comp. 2 & 1+2 \\
\hline
\multicolumn{4}{c}{\nnh}                                         \\
$N_\mathrm{col}/10^{13}\, \rm cm^{-2}$ &    $0.149 \pm 0.014$    &  $1.14 \pm 0.04$        &  $1.29\pm 0.04$   \\
$T_\mathrm{ex}/K$                      &    7.2\tablefootmark{a}     & $5.83 \pm 0.14$&     \\
$V_\mathrm{lsr}/\text{\kms}$           & $5.408 \pm 0.008$        &    $4.943 \pm 0.003 $    &     \\
$FWHM/\text{\kms}$                     &   $ 0.275 \pm 0.016 $    &    $0.474 \pm 0.012$     &     \\
\hline
\multicolumn{4}{c}{\nqn}                                         \\
$N_\mathrm{col}/10^{10}\, \rm cm^{-2}$ &   $0.44\pm 0.14$      & $1.86\pm0.19$& $2.3 \pm 0.2$     \\
$T_\mathrm{ex}/K$                      &   7.2\tablefootmark{b}     &    5.83\tablefootmark{b}      &     \\
$V_\mathrm{lsr}/\text{\kms}$           &   5.408\tablefootmark{b}      &     4.943\tablefootmark{b}    &     \\
$FWHM/\text{\kms}$                     &     0.275\tablefootmark{b}    &     0.474\tablefootmark{b}    &    \\
\hline
\end{tabular}
\tablefoot{
\tablefoottext{a}{Parameter fixed to the best-fit value found with the one-component analysis.} \\
\tablefoottext{b}{Parameter fixed to the respective one in \nnh analysis.}
}
\end{table*}

The recorded \nnh (1-0) line towards L1455 A1 presents several spectral features, as visible in Figure \ref{spec_1455}. The most evident one is found shifted by $\approx +0.9\,$\kms with respect to the main component and it is approximately six times weaker. Since it is present in all three hyperfine groups, it is most likely a second velocity component along the line of sight. Its signal to noise ratio is however insufficient to model it independently, and it is undetected in the \nqn (1-0) transition. We therefore decided to focus on the main, brighter feature only. Similarly to the L1448 C case, this also presents a wing feature on the red side of the spectrum.  \par
As for L1448 C, the fit routine is not able to converge if all parameters of the two components are unconstrained, and we therefore fix one $T_\mathrm{ex, 1}$ value to the best-fit obtained with a single velocity component. The best fit of the \nnh (1-0) line is presented in the top panels of Figure \ref{fig:L1455_2}. The model is now able to reproduce the hyperfine main beam temperatures within $15-20$\%. The two velocity components are separated in velocity by $\approx 0.5\,$\kms. The weaker one, with a column density one order of magnitude lower than the stronger component, is functional to reproduce the broad wing on the red side of the line. The fit of the \nqn (1-0) line is done with the same approach illustrated for L1448 C. The only free parameters are the column densities of the two components. The results are shown in Figure \ref{fig:L1455_2}. The best-fit values for both isotopologues are summarised in Table \ref{tab:L1455_2}. \par
Unlike for L1448 C, the total column densities of \nnh and \nqn derived with the two-component approach are consistent within uncertainties with the ones computed using the simpler, one velocity component method. As a consequence, also the derived isotopic ratio ($\text{\ratio}|_\mathrm{2c} = 560 \pm 50$) is consistent with the one presented in Sec. \ref{sec:SingleVel}.

\section{Discussion\label{sec:Discussion}}
Since the multiple-component analysis of L1448 C and L1455 A1 yields isotopic ratios consistent  within uncertainties with the ones coming from the single-component analysis, we focus on the results of the latter method (see Sec. \ref{sec:SingleVel} and Table \ref{para_results}). As already mentioned, the kinematic parameters ($V_\mathrm{lsr}$ and $FWHM$) of the \nnh and \nqn lines are always consistent within $3\sigma$ for each object, supporting the assumption that the emission from the two isotopologues arises from the same spatial region. \par

The derived excitation temperatures are in the range $5-7\, \rm K$. These values are significantly lower than the observed dust temperatures ($23-50\, \rm K$, see Table \ref{sources}). These transitions are in fact subthermally excited, so that their \tex is lower than the local gas kinetic temperature. Furthermore, the \tdust values were derived fitting continuum data at far infrared wavelengths ($\approx 60-1000\,\rm \mu m$), which are more sensitive to the warm/hot component of the dust envelope. Diazenylium, on the contrary, is expected to be destroyed by CO in the innermost part of the protostellar cores, and therefore traces preferentially a colder gas component. The large IRAM beam at 3mm, in addition, makes our observations more sensitive to the lower density envelope. \par

The last column of Table \ref{para_results} reports the derived \ratio values. The associated uncertainties are computed via standard error propagation from the column densities values. They are dominated by the uncertainties on \ncolq, which are $\approx 5-15$\%, computed in the assumption that $T_\mathrm{ex} \mathrm{(N^{15}NH^+)} = T_\mathrm{ex} \mathrm{(N_2H^+)}$. Among the six observed objects, four of them present isotopic ratios consistent with or lower than the elemental value $\text{\ratio} = 440$. Two protostars show instead fractionation ratios higher than the elemental value, namely L1448 IRS2 ($3.5\sigma$ discrepancy) and L1455 ($2.1\sigma$). The weighted average across the whole sample is $\text{\ratio}|_\mathrm{pro} = 420\pm 15$. \par
    \begin{figure}[h]
   \centering
  \includegraphics[width=0.5 \textwidth]{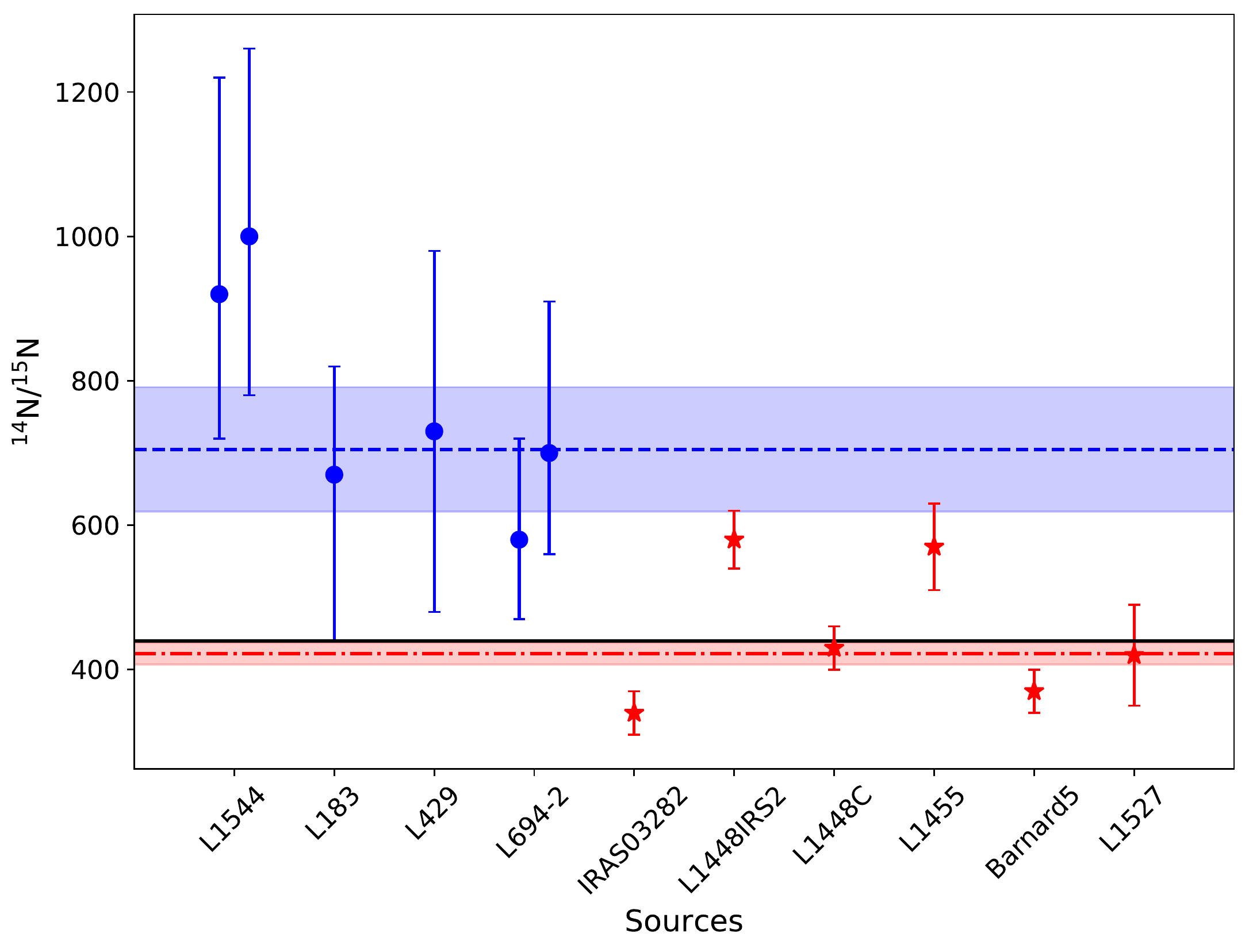}
      \caption{Nitrogen isotopic ratios in \nnh obtained in this work (red stars) compared with the values obtained towards prestellar cores (blue dots, \citealt{Redaelli18}).  Note that L1544 and L694-2 present two values, because in those sources both \nqn and \qnn were detected. The blue dashed curve shows the weighted average of the prestellar sample, and the shaded area the associated uncertainty. The red dot-dashed curve is the average for the protostellar objects. The black, solid line shows the elemental value of $\text{\ratio} = 440$. \label{fig_ratios}}
   \end{figure}
   
In Figure \ref{fig_ratios} we compare the just obtained measurements of \ratio with the observations towards prestellar cores from \cite{Redaelli18}. It is important to notice that the prestellar sample was analysed using a different approach, i.e. a fully non-LTE radiative transfer analysis. This explains why the uncertainties on the prestellar isotopic ratios are significantly larger then the protostellar ones (30\% versus 10\% on average). The errors reported in \cite{Redaelli18} represent confidence ranges, whilst the uncertainties evaluated in this work are $1\sigma$ statistical errors. \par
The nitrogen isotopic ratios measured in the prestellar sample are systematically higher than in the protostellar one. The weighted average\footnote{In order to compute the weighted average on data points with asymmetric uncertainties, we conservatively assigned to each value the highest of the two errors.} is $\text{\ratio}|_\mathrm{pre} = 700\pm90$, which is significantly higher than $\text{\ratio}|_\mathrm{pro}$. \par
\begin{figure*}[!h]
   \centering
  \includegraphics[width=.8 \textwidth]{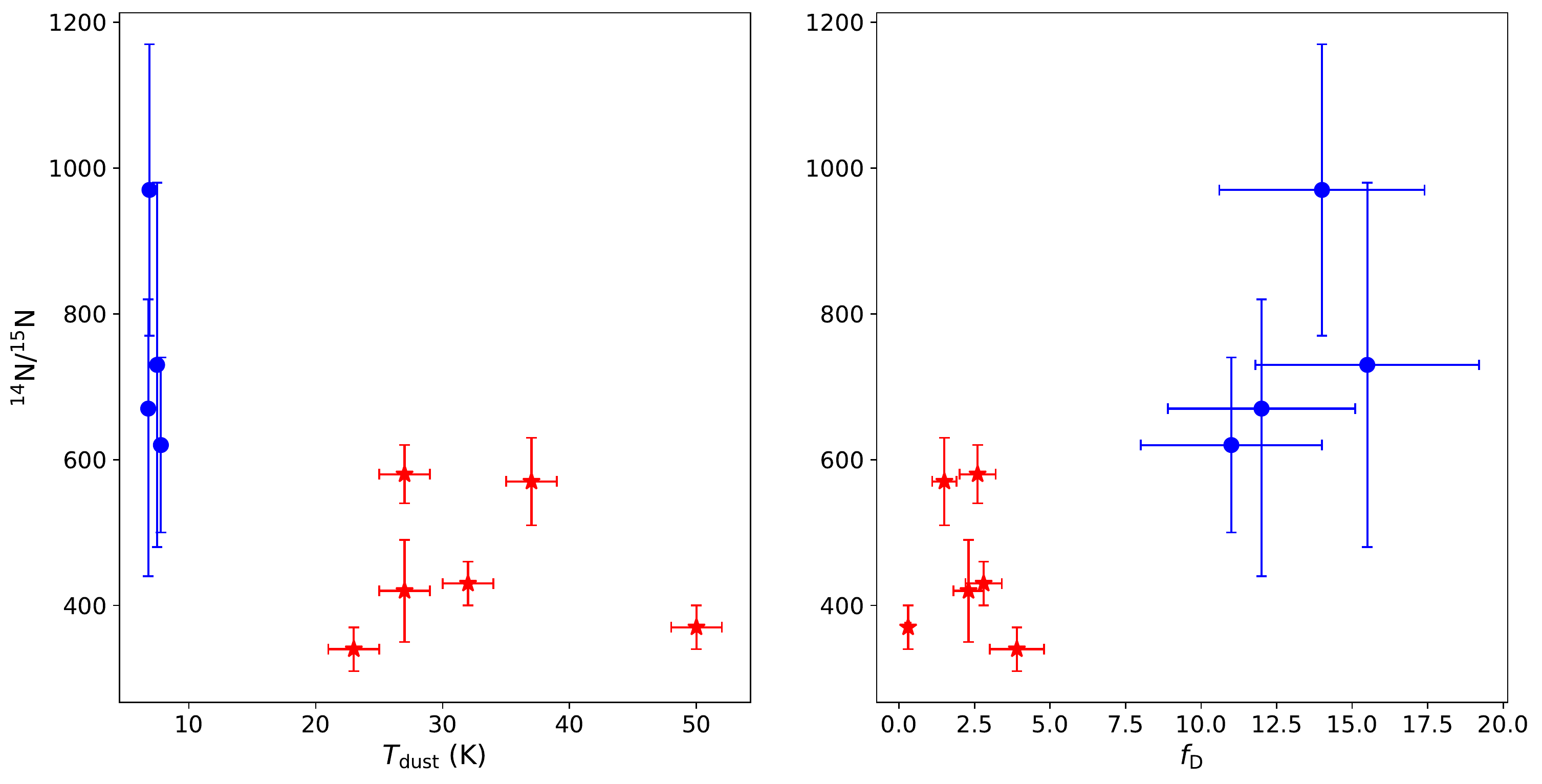}
      \caption{Scatterplots of the nitrogen isotopic ratio as a function of \tdust (left panel) and CO depletion factor (right panel). {Red stars represent protostellar cores from this work, whilst the blue dots represent the prestellar source analysed by \cite{Redaelli18}. \tdust and $f_\mathrm{D}$ values for YSOs are taken from \cite{Emprechtinger09}.  For prestellar objects, the \tdust values are taken from \cite{Chacon-Tanarro19} and \cite{Redaelli18}; the values of $f_\mathrm{D}$ are taken from \cite{Crapsi05}.} \label{fig_scatterplots}}
   \end{figure*}
\begin{figure}[h]
   \centering
  \includegraphics[width=0.45 \textwidth]{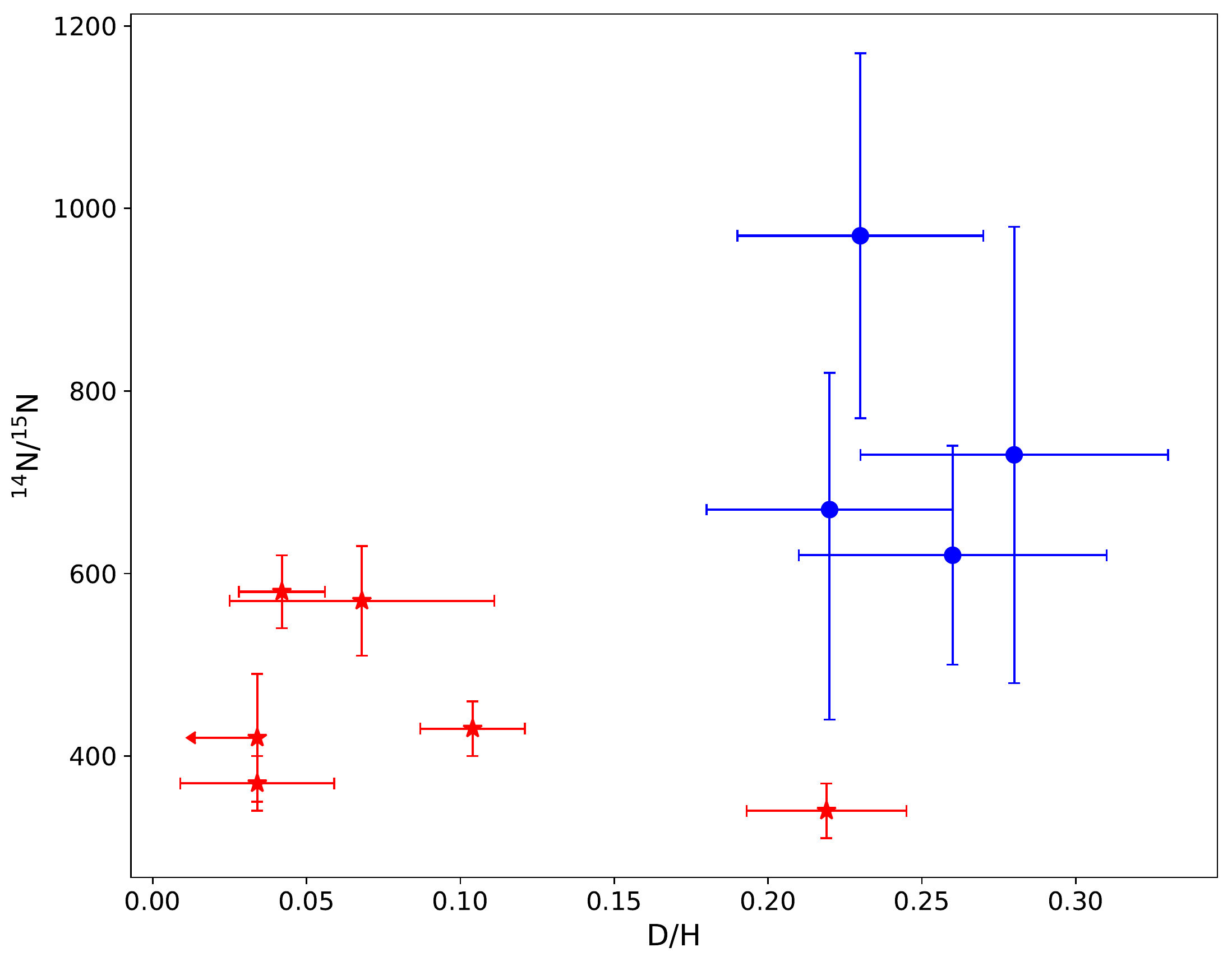}
      \caption{The correlation between nitrogen and hydrogen isotopic ratios in the six {protostellar cores} here investigated {(red stars) and in the prestellar sample of \cite{Redaelli18} (blue dots). The D/H ratio of prestellar cores come from \cite{Crapsi05}, whilst the protostellar ones from \cite{Emprechtinger09}. } \label{fig_dh}}
   \end{figure}

In the theory of \cite{Loison19}, as mentioned in the introduction, the main parameter that influences the \ratio of \nnh is the CO abundance. When CO is heavily depleted in the gas phase due to freeze-out onto the dust grains, \nqn is selectively destroyed by reaction with free electrons and, as a consequence, \ratio increases. Protostellar cores, being warmer than prestellar ones, are expected to present then lower \ratio, since CO starts to evaporate back into the gas phase at a temperature of $\approx 20\, \rm K$. {In Figure \ref{fig_scatterplots} we show the relation of the nitrogen isotopic ratios with the dust temperature (left panel)  and the CO depletion factor (right panel). The values of \tdust and $f_\mathrm{D}$ for the protostellar sample are taken from \cite{Emprechtinger09}. Concerning the prestellar cores, we report the central dust temperatures from \cite{Redaelli18} for L183, L694-2, and L429, whilst for L1544 we use the value indicated by \cite{Chacon-Tanarro19}. Since these \tdust values come from modelled profiles, they are shown without errorbars. The CO depletion factors are taken from \cite{Crapsi05}. In Figure \ref{fig_scatterplots} (and in Figure \ref{fig_dh}), for those sources where both \nqn and \qnn were observed, we report the weighted average of the two values. 
As expected, prestellar cores are significantly colder ($\text{\tdust} < 10\, \rm K$) than protostellar ones, and they also show the highest values of CO depletion ($f_\mathrm{D} > 10$).  Our data seem hence to confirm the hypothesis of \cite{Loison19} and \cite{Hily-Blant20}. }
 \par 
{Basing on the data provided by \cite{Emprechtinger09}, we can test the correlation between \tdust (and \fco)}  and \ratio also within our protostar sample. No clear trend is visible. {Furthermore}, if one excludes from the analysis Barnard 5, which is also the only Class I object in our sample, the correlation appears opposite to the expected one: the higher \tdust (and hence the lower $f_\mathrm{D}$), the higher the isotopic ratio. We fit a linear relation to both pairs of datasets, finding:
 \begin{align}
 \text{\ratio}|_\mathrm{pro} & = (10 \pm 8) \times T_\mathrm{dust} +(130 \pm 250) \; , \\ 
  \text{\ratio}|_\mathrm{pro} & = (-110 \pm 40) \times f_\mathrm{D} +(770 \pm 130) \; .
 \end{align}
 The Pearson correlation coefficients of $p_1 = 0.60$ and $p_2 = -0.75$, respectively.  Even excluding the "outlier" Barnard 5, hence, the correlation between $ \text{\ratio}|_\mathrm{pro}$ and \tdust is poor, whilst the one between $ \text{\ratio}|_\mathrm{pro}$ and $f_\mathrm{D}$ is more significant. However the correlation coefficient decreases to $p_2 = -0.13$ if the whole sample is considered. \par 
 In order to assess if there is a correlation among these parameters, and which one holds, better data are needed. In particular, \tdust and $f_\mathrm{D}$ values are derived using observations from different telescopes and hence distinct spatial resolutions. Furthermore, the far-infrared data used to estimate \tdust are known to be sensitive to the warmer component of the interstellar medium, whereas \nnh could trace the whole warm-to-cold envelope surrounding the young stellar objects. \par

We also look for a correlation between the nitrogen isotopic ratio and the hydrogen one in \nnh, using the results of \cite{Emprechtinger09} and \cite{Crapsi05}. The data are shown in Figure \ref{fig_dh}. {It is clear that in prestellar cores not only the \ratio ratio, but also the D/H one is higher than in protostellar ones. This is a well known chemical effect, due to the fact that deuteration processes are very effective at high densities and low temperatures (see \citealt{Ceccarelli14}, and references therein).}

\par
{Focusing on the protostellar cores only, a}  tentative trend is seen, in the sense that the higher the deuteration level, the lower the nitrogen isotopic ratio. A linear fit to the data, excluding L1527, for which only an upper limit in D/H is known, yields:

 \begin{equation}
 \text{\ratio}|_\mathrm{pro}  = (-455 \pm 480) \times \mathrm{D/H} +(450 \pm 60) \;,
  \end{equation}
with a Pearson correlation coefficient of $p_3 = -0.56$. This correlation is opposite to the one found by \cite{Fontani15} in high mass cores, and also to what is expected from \cite{Loison19}. In fact, the deuteration level is sensitive to the CO depletion, since carbon monoxide can effectively destroy $\rm H_2D^+$, the precursor of all deuterated gas species. One would therefore expect that when the D/H ratio is high (and hence CO is depleted from the gas phase),  \ratio also shows high values. However, we stress that the correlation that we found is weak and only tentative, and we do not draw conclusions based on it, until further data are available.

\section{Conclusions and summary}
In this work we have observed the nitrogen isotopic ratio of diazenylium in the first sample of low-mass {protostellar cores}. We detected \nnh and \nqn (1-0) lines above the $3\sigma $ level in all six targeted sources. We analysed the observations using a custom code that implements a constant \tex fit of the spectra. As illustrated in Appendix \ref{app:LTE} we do not expect excitation anomalies to be important in {protostellar cores}, as opposed to prestellar {ones}. For two sources, a two-component analysis yields better results than the single-component fit. However, the \ratio values obtained with the two approaches are consistent within the errors, and we therefore discuss the results obtained with the one-component method. \par
The weighted average of the isotopic ratio is $\text{\ratio}|_\mathrm{pro} = 420\pm 15$, which is consistent within $2\sigma$ with the protosolar value of $440$. On the contrary, this result is significantly lower than the prestellar values that are in the range $580-1000$. 
These results seem to confirm the theory of \cite{Loison19}, according to which when CO is depleted the dominant destruction pathway of diazenylium isotopologues is through dissociative recombination, and that the DR rate for \nnh is lower than that of \nqn. This would have profound implications concerning using diazenylium to trace molecular nitrogen, since it means that in cold gas the nitrogen isotopic ratio of \nnh and $\rm N_2$ are not equal.
\par
We tried to verify if a correlation between \tdust and $f_\mathrm{D}$ with \ratio is present within the YSOs sample. We do not find significant correlations, and in fact a weak trend opposite to the expected one is seen. We speculate that better data, tracing exactly the gas emitting the diazenylium lines, are needed to constrain reliably these correlations. For most sources (all but L1448 IRS2), estimations of the kinetic temperature from ammonia (1,1) and (2,2) transitions can be found in the literature \citep{Jijina99, Hatchell03}. These values are in the range $10-15\, \rm K$, hence always higher than the {derived} \tex, supporting the hypothesis of subthermal excitation for the analysed transitions. Similarly to the \tdust case, we do not find significant trend of the nitrogen isotopic ratio with the kinetic temperature. We highlight the need of a complete and coherent sample of ammonia observations, comprising also the (3,3) transition (which is needed to correctly determine the temperature of warm/hot gas), in order to further investigate this point. \par
Our results provide new, valuable inputs for chemical modellers and also for laboratory studies. In particular, we stress that further measurements of the DR rates of \nnh and \nqn in interstellar conditions are needed. These laboratory results would in fact provide definitive proof for the theory of \cite{Loison19} {and of \cite{Hily-Blant20}}. We also plan to use the data here presented as a starting point for further investigation at higher spatial resolution using for instance the NOEMA interferometer. This would give us the chance to test how the isotopic ratio varies with temperature and density within the protostellar envelope. So far, such study has been performed only in one high-mass star forming region by \cite{Colzi19}. Since low-mass star forming regions are on average closer, we could have the resolution to really unveil the role of CO freeze out/desorption in driving the nitrogen isotopic ratio.


\bibliographystyle{aa}
\bibliography{../Isubmission/Literature.bib}

\begin{thebibliography}{46}
\expandafter\ifx\csname natexlab\endcsname\relax\def\natexlab#1{#1}\fi

\bibitem[{{Altwegg} {et~al.}(2019){Altwegg}, {Balsiger}, \&
  {Fuselier}}]{Altwegg19}
{Altwegg}, K., {Balsiger}, H., \& {Fuselier}, S.~A. 2019, \araa, 57, 113

\bibitem[{{Bachiller} {et~al.}(1990){Bachiller}, {Cernicharo},
  {Martin-Pintado}, {Tafalla}, \& {Lazareff}}]{Bachiller90}
{Bachiller}, R., {Cernicharo}, J., {Martin-Pintado}, J., {Tafalla}, M., \&
  {Lazareff}, B. 1990, \aap, 231, 174

\bibitem[{{Bergin} {et~al.}(2002){Bergin}, {Alves}, {Huard}, \&
  {Lada}}]{Bergin02}
{Bergin}, E.~A., {Alves}, J., {Huard}, T., \& {Lada}, C.~J. 2002, \apjl, 570,
  L101

\bibitem[{{Bizzocchi} {et~al.}(2013){Bizzocchi}, {Caselli}, {Leonardo}, \&
  {Dore}}]{Bizzocchi13}
{Bizzocchi}, L., {Caselli}, P., {Leonardo}, E., \& {Dore}, L. 2013, \aap, 555,
  A109

\bibitem[{{Bonal} {et~al.}(2010){Bonal}, {Huss}, {Krot}, {Nagashima}, {Ishii},
  \& {Bradley}}]{Bonal10}
{Bonal}, L., {Huss}, G.~R., {Krot}, A.~N., {et~al.} 2010, \gca, 74, 6590

\bibitem[{{Caselli} {et~al.}(1995){Caselli}, {Myers}, \&
  {Thaddeus}}]{Caselli95}
{Caselli}, P., {Myers}, P.~C., \& {Thaddeus}, P. 1995, \apjl, 455, L77

\bibitem[{{Caselli} {et~al.}(2002){Caselli}, {Walmsley}, {Zucconi}, {Tafalla},
  {Dore}, \& {Myers}}]{Caselli02b}
{Caselli}, P., {Walmsley}, C.~M., {Zucconi}, A., {et~al.} 2002, \apj, 565, 344

\bibitem[{{Cazzoli} {et~al.}(2012){Cazzoli}, {Cludi}, {Buffa}, \&
  {Puzzarini}}]{Cazzoli12}
{Cazzoli}, G., {Cludi}, L., {Buffa}, G., \& {Puzzarini}, C. 2012, \apjs, 203,
  11

\bibitem[{{Ceccarelli} {et~al.}(2014){Ceccarelli}, {Caselli},
  {Bockel{\'e}e-Morvan}, {Mousis}, {Pizzarello}, {Robert}, \&
  {Semenov}}]{Ceccarelli14}
{Ceccarelli}, C., {Caselli}, P., {Bockel{\'e}e-Morvan}, D., {et~al.} 2014, in
  Protostars and Planets VI, ed. H.~{Beuther}, R.~S. {Klessen}, C.~P.
  {Dullemond}, \& T.~{Henning}, 859

\bibitem[{{Chac{\'o}n-Tanarro} {et~al.}(2019){Chac{\'o}n-Tanarro}, {Pineda},
  {Caselli}, {Bizzocchi}, {Gutermuth}, {Mason}, {G{\'o}mez-Ruiz}, {Harju},
  {Devlin}, {Dicker}, {Mroczkowski}, {Romero}, {Sievers}, {Stanchfield},
  {Offner}, \& {S{\'a}nchez-Arg{\"u}elles}}]{Chacon-Tanarro19}
{Chac{\'o}n-Tanarro}, A., {Pineda}, J.~E., {Caselli}, P., {et~al.} 2019, \aap,
  623, A118

\bibitem[{{Colzi} {et~al.}(2018){Colzi}, {Fontani}, {Caselli}, {Ceccarelli},
  {Hily-Blant}, \& {Bizzocchi}}]{Colzi18a}
{Colzi}, L., {Fontani}, F., {Caselli}, P., {et~al.} 2018, \aap, 609

\bibitem[{{Colzi} {et~al.}(2019){Colzi}, {Fontani}, {Caselli}, {Leurini},
  {Bizzocchi}, \& {Quaia}}]{Colzi19}
{Colzi}, L., {Fontani}, F., {Caselli}, P., {et~al.} 2019, \mnras, 485, 5543

\bibitem[{{Colzi} {et~al.}(2020){Colzi}, {Sipil{\"a}}, {Roueff}, {Caselli}, \&
  {Fontani}}]{Colzi20}
{Colzi}, L., {Sipil{\"a}}, O., {Roueff}, E., {Caselli}, P., \& {Fontani}, F.
  2020, arXiv e-prints, arXiv:2006.03362

\bibitem[{{Crapsi} {et~al.}(2005){Crapsi}, {Caselli}, {Walmsley}, {Myers},
  {Tafalla}, {Lee}, \& {Bourke}}]{Crapsi05}
{Crapsi}, A., {Caselli}, P., {Walmsley}, C.~M., {et~al.} 2005, \apj, 619, 379

\bibitem[{{Dalgarno} \& {Lepp}(1984)}]{Dalgarno84}
{Dalgarno}, A. \& {Lepp}, S. 1984, \apjl, 287, L47

\bibitem[{{Daniel} {et~al.}(2006){Daniel}, {Cernicharo}, \&
  {Dubernet}}]{Daniel06}
{Daniel}, F., {Cernicharo}, J., \& {Dubernet}, M.-L. 2006, \apj, 648, 461

\bibitem[{{Daniel} {et~al.}(2013){Daniel}, {G{\'e}rin}, {Roueff}, {Cernicharo},
  {Marcelino}, {Lique}, {Lis}, {Teyssier}, {Biver}, \&
  {Bockel{\'e}e-Morvan}}]{Daniel13}
{Daniel}, F., {G{\'e}rin}, M., {Roueff}, E., {et~al.} 2013, \aap, 560, A3

\bibitem[{{Dore} {et~al.}(2009){Dore}, {Bizzocchi}, {Degli Esposti}, \&
  {Tinti}}]{Dore09}
{Dore}, L., {Bizzocchi}, L., {Degli Esposti}, C., \& {Tinti}, F. 2009, \aap,
  496, 275

\bibitem[{{Emprechtinger} {et~al.}(2009){Emprechtinger}, {Caselli}, {Volgenau},
  {Stutzki}, \& {Wiedner}}]{Emprechtinger09}
{Emprechtinger}, M., {Caselli}, P., {Volgenau}, N.~H., {Stutzki}, J., \&
  {Wiedner}, M.~C. 2009, \aap, 493, 89

\bibitem[{{Fontani} {et~al.}(2015){Fontani}, {Caselli}, {Palau}, {Bizzocchi},
  \& {Ceccarelli}}]{Fontani15}
{Fontani}, F., {Caselli}, P., {Palau}, A., {Bizzocchi}, L., \& {Ceccarelli}, C.
  2015, \apjl, 808, L46

\bibitem[{{Fouchet} {et~al.}(2004){Fouchet}, {Irwin}, {Parrish}, {Calcutt},
  {Taylor}, {Nixon}, \& {Owen}}]{Fouchet04}
{Fouchet}, T., {Irwin}, P.~G.~J., {Parrish}, P., {et~al.} 2004, \icarus, 172,
  50

\bibitem[{{Frerking} {et~al.}(1982){Frerking}, {Langer}, \&
  {Wilson}}]{Frerking82}
{Frerking}, M.~A., {Langer}, W.~D., \& {Wilson}, R.~W. 1982, \apj, 262, 590

\bibitem[{{Furuya} \& {Aikawa}(2018)}]{Furuya18}
{Furuya}, K. \& {Aikawa}, Y. 2018, \apj, 857, 105

\bibitem[{{Gerin} {et~al.}(2015){Gerin}, {Pety}, {Fuente}, {Cernicharo},
  {Commer{\c{c}}on}, \& {Marcelino}}]{Gerin15}
{Gerin}, M., {Pety}, J., {Fuente}, A., {et~al.} 2015, \aap, 577, L2

\bibitem[{{Hatchell}(2003)}]{Hatchell03}
{Hatchell}, J. 2003, \aap, 403, L25

\bibitem[{{Hily-Blant} {et~al.}(2013{\natexlab{a}}){Hily-Blant}, {Bonal},
  {Faure}, \& {Quirico}}]{Hily-Blant13a}
{Hily-Blant}, P., {Bonal}, L., {Faure}, A., \& {Quirico}, E.
  2013{\natexlab{a}}, \icarus, 223, 582

\bibitem[{{Hily-Blant} {et~al.}(2017){Hily-Blant}, {Magalhaes}, {Kastner},
  {Faure}, {Forveille}, \& {Qi}}]{HilyBlant17}
{Hily-Blant}, P., {Magalhaes}, V., {Kastner}, J., {et~al.} 2017, \aap, 603, L6

\bibitem[{{Hily-Blant} {et~al.}(2020){Hily-Blant}, {Pineau des For{\^e}ts},
  {Faure}, \& {Flower}}]{Hily-Blant20}
{Hily-Blant}, P., {Pineau des For{\^e}ts}, G., {Faure}, A., \& {Flower}, D.~R.
  2020, arXiv e-prints, arXiv:2009.06393

\bibitem[{{Hily-Blant} {et~al.}(2013{\natexlab{b}}){Hily-Blant}, {Pineau des
  For{\^e}ts}, {Faure}, {Le Gal}, \& {Padovani}}]{Hily-Blant13b}
{Hily-Blant}, P., {Pineau des For{\^e}ts}, G., {Faure}, A., {Le Gal}, R., \&
  {Padovani}, M. 2013{\natexlab{b}}, \aap, 557, A65

\bibitem[{{Jijina} {et~al.}(1999){Jijina}, {Myers}, \& {Adams}}]{Jijina99}
{Jijina}, J., {Myers}, P.~C., \& {Adams}, F.~C. 1999, \apjs, 125, 161

\bibitem[{{Kahane} {et~al.}(2018){Kahane}, {Jaber Al-Edhari}, {Ceccarelli},
  {L{\'o}pez-Sepulcre}, {Fontani}, \& {Kama}}]{Kahane18}
{Kahane}, C., {Jaber Al-Edhari}, A., {Ceccarelli}, C., {et~al.} 2018, \apj,
  852, 130

\bibitem[{{Keto} {et~al.}(2015){Keto}, {Caselli}, \& {Rawlings}}]{Keto15}
{Keto}, E., {Caselli}, P., \& {Rawlings}, J. 2015, \mnras, 446, 3731

\bibitem[{{Lawson} {et~al.}(2011){Lawson}, {Osborne}, \& {Adams}}]{Lawson11}
{Lawson}, P.~A., {Osborne}, David, J., \& {Adams}, N.~G. 2011, International
  Journal of Mass Spectrometry, 304, 41

\bibitem[{{Loison} {et~al.}(2019){Loison}, {Wakelam}, {Gratier}, \&
  {Hickson}}]{Loison19}
{Loison}, J.-C., {Wakelam}, V., {Gratier}, P., \& {Hickson}, K.~M. 2019,
  \mnras, 484, 2747

\bibitem[{{Marty} {et~al.}(2011){Marty}, {Chaussidon}, {Wiens}, {Jurewicz}, \&
  {Burnett}}]{Marty11}
{Marty}, B., {Chaussidon}, M., {Wiens}, R.~C., {Jurewicz}, A.~J.~G., \&
  {Burnett}, D.~S. 2011, Science, 332, 1533

\bibitem[{{Melosso} {et~al.}(2020){Melosso}, {Bizzocchi}, {Sipil{\"a}},
  {Giuliano}, {Dore}, {Tamassia}, {Martin-Drumel}, {Pirali}, {Redaelli}, \&
  {Caselli}}]{Melosso20}
{Melosso}, M., {Bizzocchi}, L., {Sipil{\"a}}, O., {et~al.} 2020, arXiv
  e-prints, arXiv:2007.07504

\bibitem[{{Nier}(1950)}]{Nier50}
{Nier}, A.~O. 1950, Physical Review, 77, 789

\bibitem[{{Pagani} {et~al.}(2007){Pagani}, {Bacmann}, {Cabrit}, \&
  {Vastel}}]{Pagani07}
{Pagani}, L., {Bacmann}, A., {Cabrit}, S., \& {Vastel}, C. 2007, \aap, 467, 179

\bibitem[{{Redaelli} {et~al.}(2018){Redaelli}, {Bizzocchi}, {Caselli}, {Harju},
  {Chac{\'o}n-Tanarro}, {Leonardo}, \& {Dore}}]{Redaelli18}
{Redaelli}, E., {Bizzocchi}, L., {Caselli}, P., {et~al.} 2018, \aap, 617, A7

\bibitem[{{Redaelli} {et~al.}(2019){Redaelli}, {Bizzocchi}, {Caselli},
  {Sipil{\"a}}, {Lattanzi}, {Giuliano}, \& {Spezzano}}]{Redaelli19}
{Redaelli}, E., {Bizzocchi}, L., {Caselli}, P., {et~al.} 2019, \aap, 629, A15

\bibitem[{{Roueff} {et~al.}(2015){Roueff}, {Loison}, \& {Hickson}}]{Roueff15}
{Roueff}, E., {Loison}, J.~C., \& {Hickson}, K.~M. 2015, \aap, 576, A99

\bibitem[{{Ruaud} {et~al.}(2016){Ruaud}, {Wakelam}, \& {Hersant}}]{Ruaud16}
{Ruaud}, M., {Wakelam}, V., \& {Hersant}, F. 2016, \mnras, 459, 3756

\bibitem[{{Schlafly} {et~al.}(2014){Schlafly}, {Green}, {Finkbeiner}, {Rix},
  {Bell}, {Burgett}, {Chambers}, {Draper}, {Hodapp}, {Kaiser}, {Magnier},
  {Martin}, {Metcalfe}, {Price}, \& {Tonry}}]{Schlafly14}
{Schlafly}, E.~F., {Green}, G., {Finkbeiner}, D.~P., {et~al.} 2014, \apj, 786,
  29

\bibitem[{{Wampfler} {et~al.}(2014){Wampfler}, {J{\o}rgensen}, {Bizzarro}, \&
  {Bisschop}}]{Wampfler14}
{Wampfler}, S.~F., {J{\o}rgensen}, J.~K., {Bizzarro}, M., \& {Bisschop}, S.~E.
  2014, \aap, 572, A24

\bibitem[{{Wirstr{\"o}m} \& {Charnley}(2018)}]{Wirstrom17}
{Wirstr{\"o}m}, E.~S. \& {Charnley}, S.~B. 2018, \mnras, 474, 3720

\bibitem[{{Zucker} {et~al.}(2018){Zucker}, {Schlafly}, {Speagle}, {Green},
  {Portillo}, {Finkbeiner}, \& {Goodman}}]{Zucker18}
{Zucker}, C., {Schlafly}, E.~F., {Speagle}, J.~S., {et~al.} 2018, \apj, 869, 83

\end{thebibliography}

\appendix

\section{On the constant \tex ($C\_\text{\tex}$) assumption\label{app:LTE}}
Strictly speaking, a local thermodynamic equilibrium (LTE) analysis assumes that the excitation temperature of all the hyperfine components in all the rotational lines is the same, and that it is equal to the gas kinetic temperature $T_\mathrm{K}$. In our approach, however, we assume that there is one \tex value shared among all the hyperfine components of the (1-0) transition, which may not coincide with $T_\mathrm{K}$. This assumption, known as $C\_\text{\tex}$, is more relaxed than LTE, but it still needs justification. \cite{Daniel06} extensively studied the problem of anomalies in the hyperfine intensity ratios of \nnh, originally reported by \cite{Caselli95}, and found that these effects are more severe at very high volume densities, low temperatures, and high optical depths. These are the physical conditions found in evolved prestellar cores, where $n> 10^5\, \rm cm^{-3}$ and $T<10\,$K. This justified the choice made by the authors in \cite{Redaelli18} to implement a fully non-LTE radiative transfer analysis to model \nnh and \nqn in the sample of prestellar cores.  \par
    \begin{figure*}[!b]
   \centering
  \includegraphics[width=0.8 \textwidth]{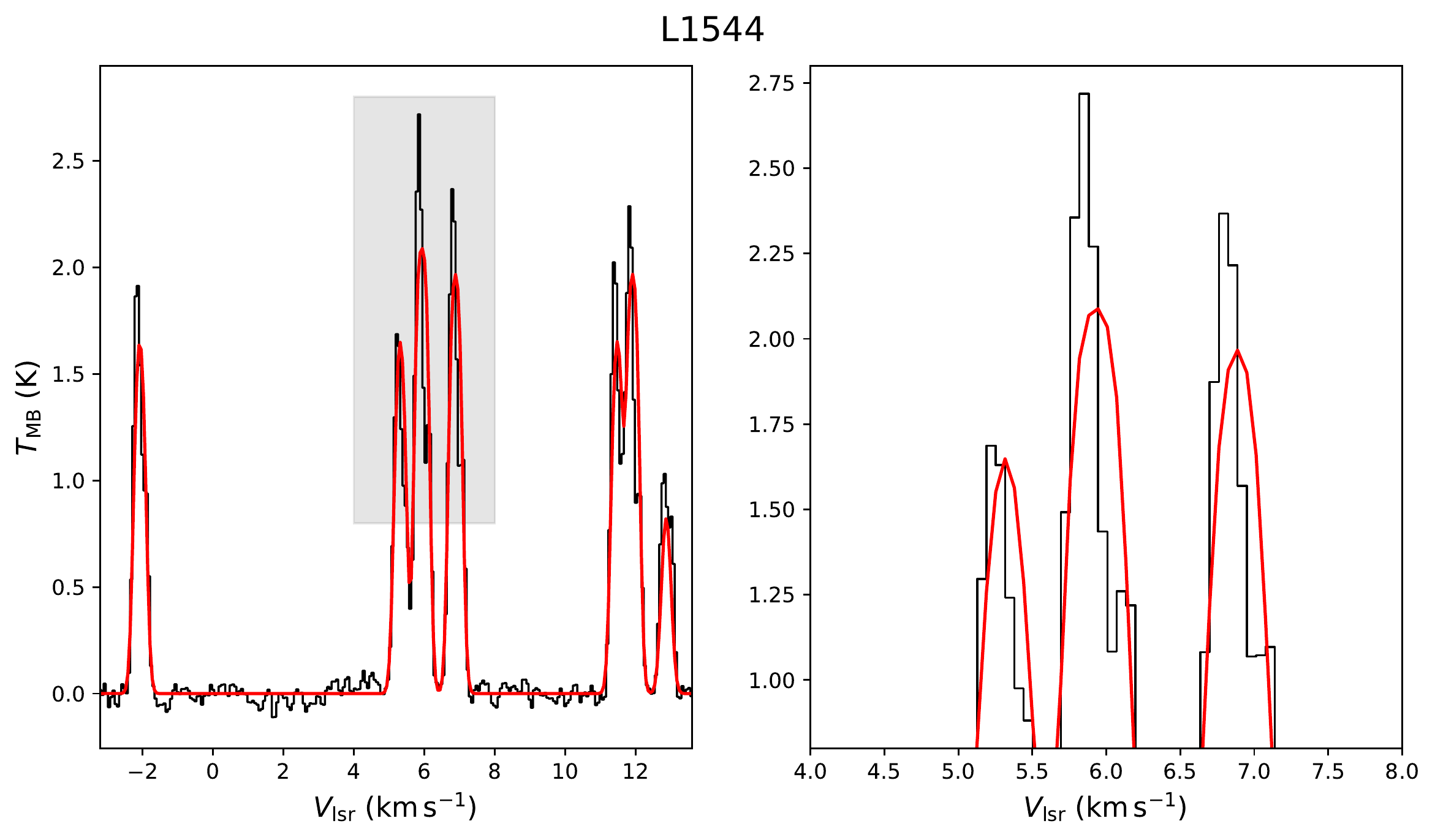}
      \caption{Modelling of \nnh (1-0) trnasitions in the prestellar core L1544. The data, taken from \cite{Redaelli18}, were observed with the same instrumental setup presented in Sec. \ref{sec:observatios}. The right panel shows the zoom-in of the grey-shaded area in the left one, highlighting the problem with the hyperfine intensities estimation in the central triplet. \label{fig:L1544}}
   \end{figure*}
Figure \ref{fig:L1544}  shows the results of the fit with our custom code on the \nnh (1-0) transition in L1544, a very evolved prestellar core. The routine is unable to reproduce the different hyperfine components intensities. In particular, with the exception of two components, the intensities are underestimated by $20-25$\%. As a result, the column density and/or the excitation temperature is underestimated, since the total flux is not reproduced. This translates directly in unreliable estimations of \ncol, and also of \ncolq (since this is estimated using the \tex value of the main isotopologue). \par
In protostellar {cores}, however, conditions are different, and they make the excitation anomaly effects less critical. Due to higher temperature and more turbulent motions, lines are in general broader. The average linewidth of the sample presented in \cite{Redaelli18} is $<FWHM>_\mathrm{pre}= 0.3\,$\kms, whilst it is $<FWHM>_\mathrm{pro}= 0.5 \,$\kms for the YSOs here analysed.  Since the optical depth is inversely proportional to the line width (see e.g. Eq. 3 in \citealt{Redaelli19}), this means that {for protostellar cores the total optical depth $\tau_\mathrm{tot}$, summed over all the hyperfine components,}  is lower than for prestellar {cores}. Our code derives also the line total optical depths, which have a mean value for the YSOs of $<\tau_\mathrm{tot}>_\mathrm{pro} \approx 7.0$. In comparison, the optical depths in the prestellar sample are always $>13.0$, and their average is $<\tau_\mathrm{tot}>_\mathrm{pre} \approx 16$. Furthermore, broader lines means that selective trapping effects, which contribute to hyperfine anomalies, are less severe. We therefore expect that the $C\_\text{\tex}$ assumption holds better for protostellar cores than for prestellar ones. This is supported by the fact that our fitting routine is overall able to reproduce the observed spectra, especially for those sources without evidence of multiple velocity components. \par
We also want to highlight the difficulties that performing a full non-LTE, non-$C\_\text{\tex}$ analysis carries. In order to implement it, the physical structure of the source in terms of temperature, density, and kinematics is needed. Prestellar cores can easily be modelled as spherically symmetric, especially around their centre, and far-infrared data (such as from Herschel) can be used to characterise the \tdust and volume density profiles. The kinematics, usually due to infall or expansion motions, is also often one-dimensional, and can be inferred from spectroscopic data (see for instance \citealt{Keto15}). The structure of protostellar cores is on the contrary more complex. Due to the presence of warm/hot dust at the centre and a colder envelope surrounding it, many wavelenghts are needed to constrain the dust thermal emission and thus its temperature and density distribution. Furthermore, the presence of molecular outflows and accretion motions make the structure deviate strongly for a 1-D assumption. \par
In conclusion, modelling the physical structure of our sample of protostellar cores, a fundamental step for non-$C\_\text{\tex}$ analysis, require extensive datasets, available for all the targeted sources, which is beyond the scopes of this work. At the same time, for the aforementioned reasons, we do not expect significant hyperfine anomalies. We conclude that the assumption of constant \tex holds reasonably.

\end{document}